\documentclass[twocolumn]{aastex701} 

\usepackage{soul}
\usepackage[T1]{fontenc}
\usepackage[utf8]{inputenc}
\usepackage{graphicx}
\usepackage{xcolor}
\usepackage{amsmath}
\usepackage{amssymb}
\usepackage{siunitx}
\usepackage{mathptmx}
\usepackage{txfonts}
\usepackage{subcaption}
\usepackage[font=small]{caption}
\usepackage{blindtext}
\usepackage{appendix}
\usepackage{dsfont}
\usepackage{booktabs}
\usepackage{newunicodechar}
\usepackage{makecell}

\providecommand{\bjdtdb}{\ensuremath{\mathrm{BJD}_{\mathrm{TDB}}}}
\providecommand{\feh}{\ensuremath{\left[\mathrm{Fe}/\mathrm{H}\right]}}
\providecommand{\teff}{\ensuremath{T_{\mathrm{eff}}}}
\providecommand{\ecosw}{\ensuremath{e\cos{\omega_*}}}
\providecommand{\esinw}{\ensuremath{e\sin{\omega_*}}}

\providecommand{\msun}{\ensuremath{M_\odot}}
\providecommand{\rsun}{\ensuremath{R_\odot}}
\providecommand{\lsun}{\ensuremath{L_\odot}}

\providecommand{\mj}{\ensuremath{\,M_{\mathrm{J}}}}
\providecommand{\rj}{\ensuremath{\,R_{\mathrm{J}}}}

\providecommand{\me}{\ensuremath{\,M_{\mathrm{E}}}}
\providecommand{\re}{\ensuremath{\,R_{\mathrm{E}}}}

\providecommand{\fave}{\langle F \rangle}
\providecommand{\fluxcgs}{10$^9$ erg s$^{-1}$ cm$^{-2}$}

\usepackage{makecell}

\begin{document} 
\pagestyle{plain}


\title{TOI-7154b: A close-in massive brown dwarf in an eccentric orbit \footnote{Most of the work related to this paper was carried out at the Astronomy $\&$ Astrophysics Division, Physical Research Laboratory, Ahmedabad, India, including the speckle-imaging analysis with the PRL 2.5 m telescope, the PARAS-2 radial-velocity analysis, the TESS photometry analysis, and the Global modelling and fitting.}}

\author[0009-0002-4044-3597]{Rohan Ch. Das}
\affiliation{Department of Physics, Mizoram University, Aizawl-796004, India}
\affiliation{Astronomy $\&$ Astrophysics Division, Physical Research Laboratory, Ahmedabad 380009, India}
\email[]{rohanchdas@yahoo.com}

\author[0000-0002-8804-650X]{Churchil Dwivedi} 
\affiliation{Astronomy $\&$ Astrophysics Division, Physical Research Laboratory, Ahmedabad 380009, India}
\email[]{churchil@prl.res.in}  

\author[0000-0001-8983-5300]{Rishikesh Sharma}
\affiliation{Astronomy $\&$ Astrophysics Division, Physical Research Laboratory, Ahmedabad 380009, India}
\email{rishikesh@prl.res.in}  

\author[0000-0002-5181-0463]{Hareesh G. Bhaskar}
\affiliation{Department of Astronomy, Indiana University, Bloomington, IN 47405, USA}
\email{bhareeshg@gmail.com}  

\author[0009-0000-4834-5612]{K.J. Nikitha}
\affiliation{Astronomy $\&$ Astrophysics Division, Physical Research Laboratory, Ahmedabad 380009, India}
\email{nikitha@prl.res.in}  

\author[0000-0001-6637-5401]{Allyson Bieryla }
\affiliation{Center for Astrophysics | Harvard $\&$ Smithsonian, 60 Garden St., Cambridge MA 02138, US}
\email{abieryla@cfa.harvard.edu}  

\author[0000-0003-1713-3208]{Boris S. Safonov}
\affiliation{Sternberg Astronomical Institute Lomonosov Moscow State University}
\email{safonov10@gmail.com}  

\author[0009-0006-9996-1814]{Shubhendra N. Das}
\affiliation{Astronomy $\&$ Astrophysics Division, Physical Research Laboratory, Ahmedabad 380009, India}
\affiliation{Indian Institute of Technology, Gandhinagar 382355, India}
\email{shubhendra@prl.res.in}  

\author[0000-0002-3815-8407]{Abhijit Chakraborty}
\affiliation{Astronomy $\&$ Astrophysics Division, Physical Research Laboratory, Ahmedabad 380009, India}
\email[show]{abhijit@prl.res.in}  

\author[0000-0003-4228-2686]{Lalthakimi Zadeng}
\affiliation{Department of Physics, Mizoram University, Aizawl-796004, India}
\email[show]{kimizadeng@mzu.edu.in}  

\author[0000-0001-9911-7388]{David W. Latham}
\affiliation{Center for Astrophysics | Harvard $\&$ Smithsonian, 60 Garden St., Cambridge MA 02138, US}
\email{dlatham@cfa.harvard.edu}  

\author[0000-0003-0670-5821]{Neelam J.S.S.V. Prasad}
\affiliation{Astronomy $\&$ Astrophysics Division, Physical Research Laboratory, Ahmedabad 380009, India}
\email{prasadgn@prl.res.in}  

\author[0000-0003-1373-4583]{Kapil K. Bharadwaj}
\affiliation{Astronomy $\&$ Astrophysics Division, Physical Research Laboratory, Ahmedabad 380009, India}
\email{kapilb@prl.res.in}  

\author[0009-0008-4890-9527]{Kevikumar A. Lad}
\affiliation{Astronomy $\&$ Astrophysics Division, Physical Research Laboratory, Ahmedabad 380009, India}
\email{kevikumar@prl.res.in}  

\author[0009-0001-3782-4308]{Ashirbad Nayak}
\affiliation{Astronomy $\&$ Astrophysics Division, Physical Research Laboratory, Ahmedabad 380009, India}
\email{ashirbad@prl.res.in}  

\correspondingauthor{Rohan Ch. Das}
\email{rohanchdas@yahoo.com} 

\correspondingauthor{Churchil Dwivedi}
\email{churchil@prl.res.in} 

\begin{abstract}
    
We report here the discovery and characterization of a high-mass transiting brown dwarf in a close-in orbit around its host star, TOI-7154. Initially, the host star was identified as an exoplanetary candidate from the TESS photometry data. Later, with the mass measurements from the RV follow-up using the PARAS-2 and TRES spectrographs, the companion is found to be sub-stellar in nature. TOI-7154, is a G-type main-sequence metal-rich star metallicity $\mathrm{[Fe/H]} = 0.154^{+0.077}_{-0.075}\,\text{dex}$, effective temperature $T_{\mathrm{eff}} = 5564^{+100}_{-110}\,\text{K}$, mass $M_\star = 0.939^{+0.047}_{-0.043}\,M_{\odot}$, radius $R_\star = 0.949^{+0.032}_{-0.030}\,R_{\odot}$, and surface gravity $\log g = 4.456^{+0.036}_{-0.036}$. With the joint analysis of the TESS photometry and the PARAS-2 and TRES radial velocities we found that TOI-7154b orbits its host star in $P = 8.860073\pm 0.000029\,\text{d}$, eccentric ($e = 0.2482 \pm 0.0024$) orbit and its radius is smaller than that of Jupiter ($R_{b} = 0.827^{+0.040}_{-0.037}\,R_{\mathrm{J}}$). With a mass near the hydrogen-burning boundary ($M_{b} = 71.7^{+2.4}_{-2.2}\,M_{\mathrm{J}}$) which separates brown dwarfs from very low-mass stars, TOI-7154b occupies a critical position in the regime for probing the transition between sub-stellar and stellar objects. The system is very old, with its age estimated to be $7.2^{+3.9}_{-3.6}\,\text{Gyr}$ by MIST isochrones, while Galactic kinematics indicate an age of $\sim4-5\,\text{Gyr}$. {Our tidal evolution simulations indicate a stellar dissipation factor of $Q_\star'\lesssim10^6$. Since the presence of any companion is currently ruled out by observations, the presence of eccentricity in this old system is, therefore, indicative of it having stellar-like fragmentation origins.}

\end{abstract}



\section{Introduction} \label{sec:intro}

Brown Dwarfs (BDs) are sub-stellar objects with masses insufficient to sustain stable hydrogen fusion in their interiors. They are generally defined to have masses in the range $\sim13-80\,M_{\mathrm J}$, where the lower limit corresponds to the onset of deuterium burning \citep{spiegel2011} and the upper limit approximates the threshold of sustained hydrogen burning, which lies between $\sim75-80\,M_{\mathrm J}$ depending on metallicity and evolutionary models \citep{baraffe2002, dieterich2014}. These boundaries blur the distinction between massive giant planets \& low-mass BDs and the high-mass BDs \& low-mass stars \citep{Persson+2019, barkaoui2025}. BDs, therefore, occupy a mass regime in which both planetary and stellar formation mechanisms may operate, making them crucial for understanding the transition between star and planet formation. Current theoretical and observational evidences indicate that BDs do not form via a single mechanism. Instead, their formation pathways appear to depend on their masses. Studies suggest that the high-mass BDs ($\gtrsim40-50\,M_{\mathrm J}$) are formed through turbulent fragmentation of molecular clouds, analogous to the formation of low-mass stars \citep{bate2009, chabrier2014, ma2014}. In contrast, low-mass BDs may also arise from disk-based processes such as gravitational instability \citep{cameron1978, boss1997} or core accretion mechanism \citep{pollack1996, ma2014, kratter2016} as giant planets. {These formation pathways are expected to leave observable dynamical imprints, including differences in orbital eccentricities, period distributions, period–eccentricity relations, and spin–orbit obliquity \citep{ma2014, Bowler2020, Gan2025}}.

{Beyond their physical properties, the orbital properties of close-in transiting BDs can provide critical insights into their formation and dynamical histories. Disk-driven migrations typically produce low-eccentricity orbits due to strong eccentricity damping in the protoplanetary disk \citep{goldreich1980, tanaka2004}. In contrast, BDs formed through star-like cloud fragmentation at wide separations may subsequently undergo dynamical excitation via Kozai–Lidov oscillations \citep{kozai1962, lidov1962, wu2003}, planet-planet or planet-companion scattering \citep{chatterjee2008, nagasawa2008}, before migrating inward and experiencing strong tidal interactions with the primary star. As a result, moderate eccentricities ($e\sim0.1-0.4$) at short periods ($P\lesssim10\,\text{d}$) can preserve signatures of high-eccentricity migration or ongoing tidal circularisation \citep{mazeh2008, matsumura2010}.}

{Tidal interactions further shape the orbital evolution of close-in sub-stellar companions. Since BDs are dense objects, their tidal evolution can differ significantly from that of giant planets or low-mass stars \citep{ogilvie2014, barker2020}. There are a plethora of results that could be attributed to the examination of tidal interactions, that would have otherwise remained a mystery. One such example is the discrepancy between the observed and predicted radii of transiting exoplanets and brown dwarfs \citep{Jackson+2008, Jackson+2009, Henderson+2024} which can be attributed to the effect of tidal heating, the observed narrow range of eccentricity in close-in exoplanets compared to the broad range for wide-orbit systems caused by the tidal circularisation of close-in systems, and the evolution of orbital separation due to tides which eventually leads to the accretion of the planet by the host star after crossing the Roche limit \citep{Jackson+2008, Jackson+2009}. Thus, the observed eccentricities and tidal decay timescales of close-in BDs can put constraints on their formation mechanisms, past dynamical interactions, and orbital evolution.}

{One of the prominent features of the  BD population is the \textit{brown dwarf desert} -- a scarcity of BDs with short orbital periods ($P_{\text{orb}}\lesssim10-100\,\text{d}$) around FGK stars \citep{marcy2000, grether2006, sahlmann2011}. Although more than $2000$ BDs have been identified through wide-field imaging surveys \citep{johnston2015}, only $\sim50$ transiting BDs are currently known \citep[e.g.][]{Persson+2019, mireles2023, barkaoui2025}. Transiting BDs are rare, but important, as they offer a unique opportunity to directly measure mass and radius through combined photometric and high-resolution spectroscopic observations, independent of the sub-stellar evolutionary models. It is helpful in constraining the BD mass-radius relation and their subsequent evolutionary models. Despite their rarity, the synergy between space-based photometric surveys, such as CoRoT \citep{corot}, Kepler \citep{kepler}, TESS \citep{Ricker2015}, and ground-based radial velocity (RV) spectrographs, such as HARPS \citep{harps}, PARAS \citep{chakraborty2014_PARAS}, TRES \citep{gaborthesistres}, FIES \citep{feis}, {and others}, have improved the transiting BD population in recent years. TESS, in particular, provides continuous high-precision light curves for bright nearby stars, enabling the detection of transiting BD companions across a wide mass range.} 

In this work, we report the discovery and characterisation of TOI-7154b, a transiting close-in high-mass BD in an eccentric orbit detected with the TESS photometry data and confirmed with high-resolution RV observations from the PARAS-2 and the TRES spectrographs. The rest of the paper is organized as follows: Section 2 describes the photometric, spectroscopic, and speckle imaging observations; Section 3 elaborates the various analyses of the host star and the system along with the joint modeling of the light curves and RVs; and Section 4 presents the newly detected BD in the context of BD evolutionary models and the tidal evolution of the system. It also discusses the possible formation pathways of the BD. In Section 5, we provide a brief summary of our results.

\section{Observations} \label{sec:obs}
\begin{figure*}[ht!]
\centering
\includegraphics[width=0.9\paperwidth]{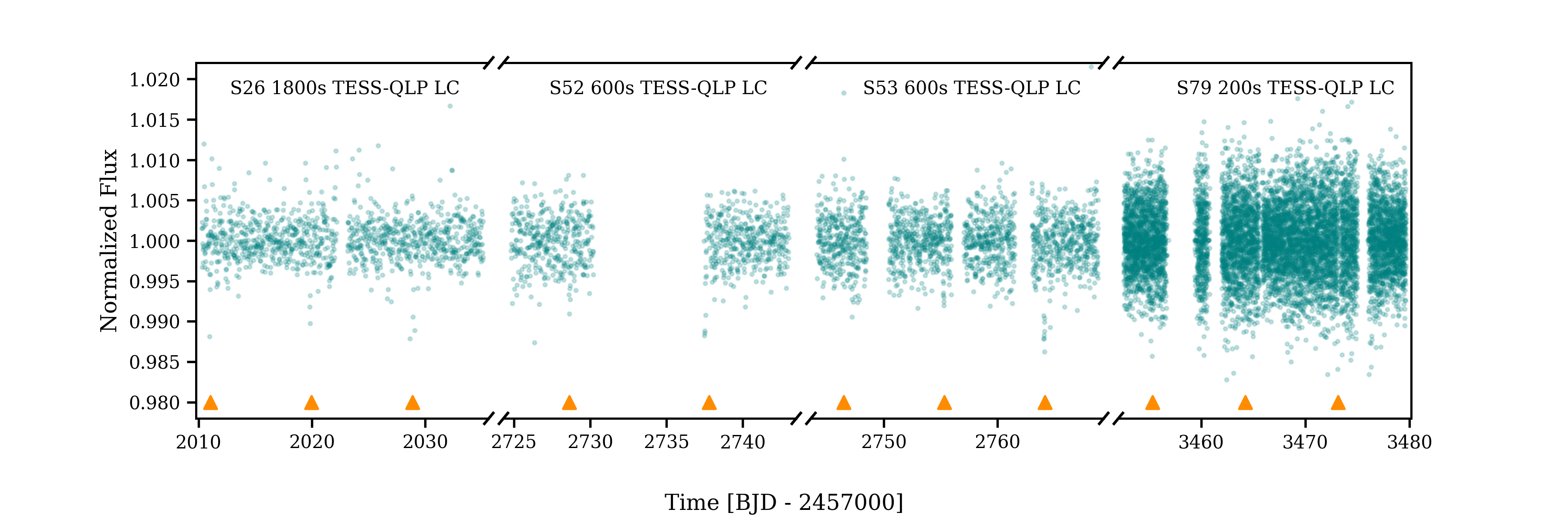}
\includegraphics[width=0.5\paperwidth]{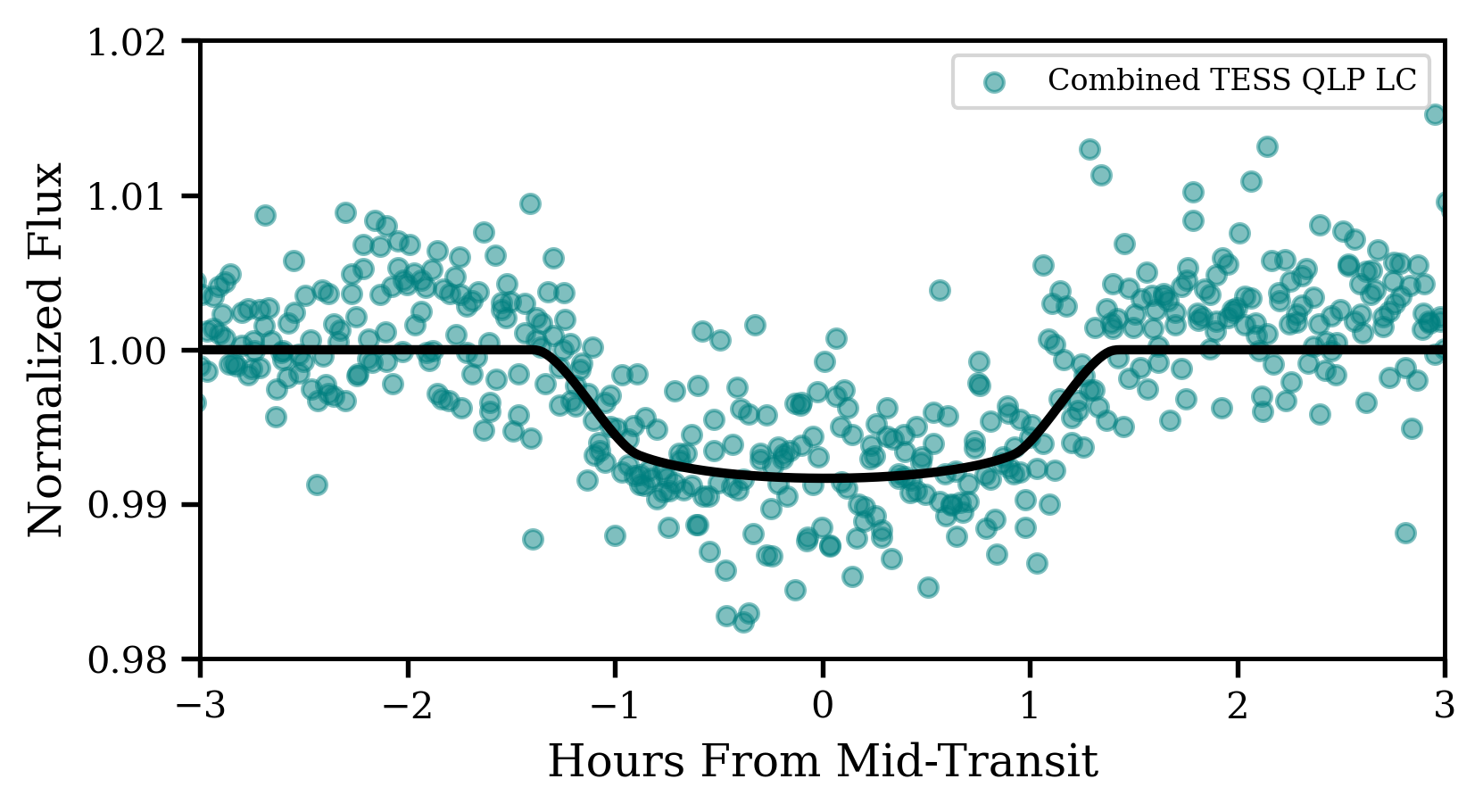}
\caption{Detrended TESS light curve (LC) from sectors 26, 52, 53, and 79 shown in teal color. It displays the full TESS LC plotted against time. {{The full LC is displayed in the upper panel, where the triangles mark the time at which transit occurs.}} The phase-folded LC from all sectors is shown in the lower panel with the best-fit transit model is overplotted in black.}
\label{fig:tesslc}
\end{figure*}
\subsection{TESS}\label{sec:tess_obs}
TOI-7154 (TIC 377622130) was observed by the \textit{Transiting Exoplanet Survey Satellite} (TESS) as part of its ongoing all-sky survey designed to detect transiting exoplanets around bright nearby stars. The system was monitored across multiple sectors and at different cadences. It was first observed at a 30-minute cadence in Sector 26 (from June 9, 2020 to July 3, 2020), followed by 600-second cadence observations in Sectors 52 and 53 (from May 19, 2022 to June 12, 200 and June 13, 2022 to July 7, 2022, respectively), and most recently at a 200-second cadence in Sector 79 (from May 5, 2024 to June 3, 2024). These time intervals provide a photometric baseline of nearly four years, allowing for a consistent investigation of both short-term and long-term photometric variability. The {\it TESS} light curve data, which is publicly available in the Mikulski Archive for Space Telescopes (MAST\footnote{\url{https://mast.stsci.edu/portal/}}) portal has been used in this study, which can be accessed with the identifier \dataset[10.17909/ergc-ph96]{http://dx.doi.org/10.17909/ergc-ph96}. The light curves have been extracted using the \textit{Quick-Look Pipeline} (QLP; \citealt{Huang2020}), which is optimised to process TESS Full Frame Image (FFI) data and deliver high-quality photometric time series for bright targets. The QLP performs aperture photometry using dynamically chosen apertures that minimise background contamination and systematic noise. For our analysis, we used the \texttt{KSPSAP} (or \texttt{DET}) flux produced by QLP for transit characterization, as it provides detrended fluxes corrected for common instrumental and background effects. We removed all cadences flagged as poor quality by excluding data points with non-zero TESS \texttt{QUALITY} bits that indicate instrumental events, scattered light contamination, or QLP-specific flags as mentioned in \citet{Huang2020} and timely updated at the TESS QLP webpage\footnote{\url{https://tess.mit.edu/qlp/}}. Furthermore, we normalized the flux around its median value and applied a {second-order polynomial fit} to remove long-term trends while masking all the transit events{, where each sector is fit individually for handling the gaps in the data}. This step ensured a uniform photometric baseline across all sectors and cadences. The cleaned and detrended light curves were subsequently combined and used for transit modeling (see Section~\ref{sec:global_modeling}). The normalized TESS light curves are plotted in Figure~\ref{fig:tesslc}, along with the phase-folded light curve.

\subsection{Speckle observations}
\subsubsection{Speckle observations with PRL 2.5-m Telescope}
\begin{figure}[t!]
\centering
\includegraphics[width=0.88\columnwidth]{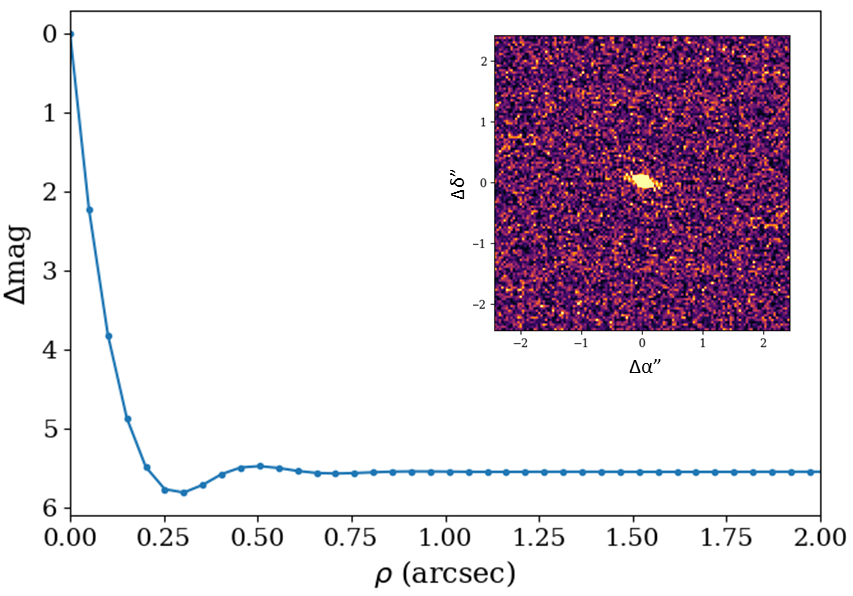}
\hspace{0.8cm}
\includegraphics[width=\columnwidth]{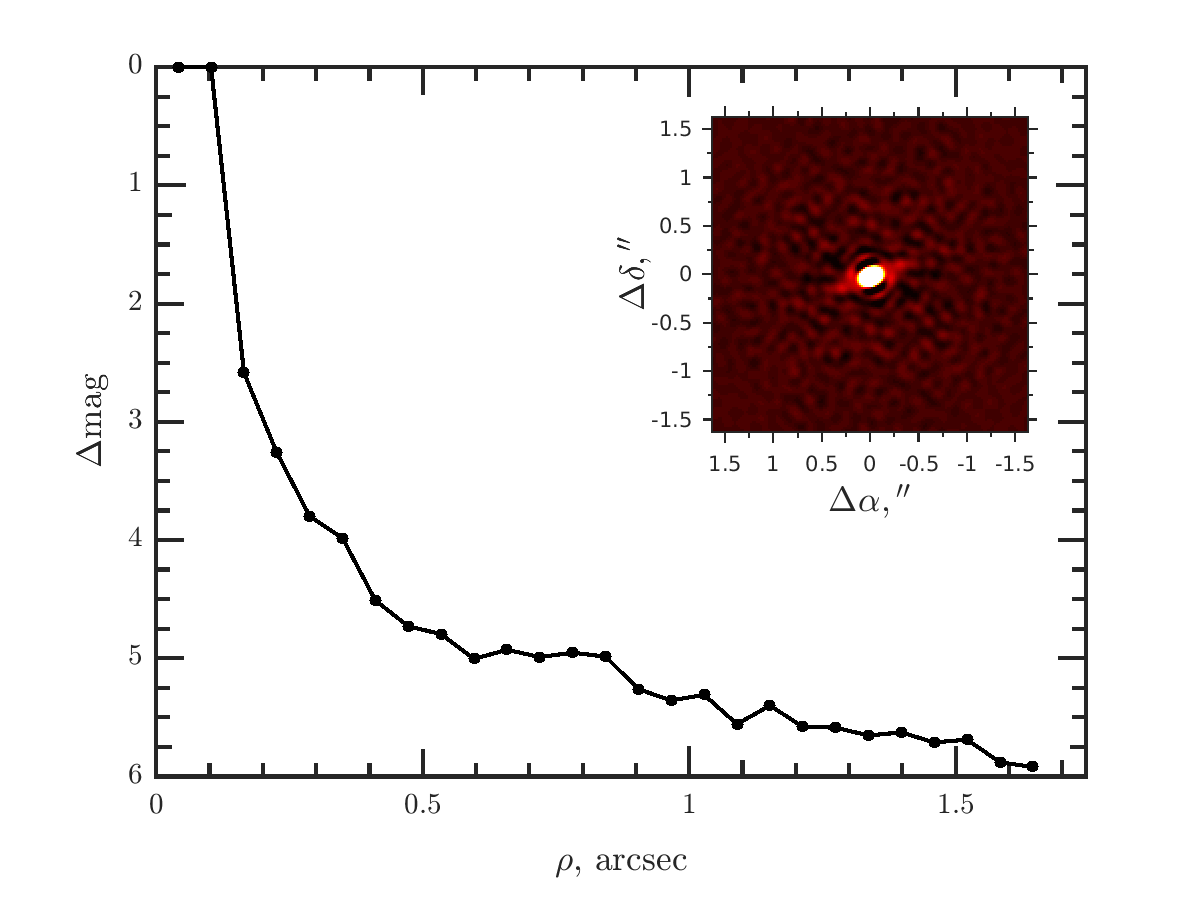}
\caption{Contrast curve in $V$ band for TOI-7154 obtained from the speckle imager for PRL 2.5m telescope (top panel). The contrast curve from {{SPeckle Polarimeter}} (SPP) speckle analysis in the $I_c$ band (bottom panel). The speckle ACF is displayed as an inset. No stellar companions are detected.}
\label{fig:speckle}
\end{figure}

Speckle imaging observations of TOI-7154 were conducted in April 2025 using the Speckle Imager \citep{prl2.5m_and_paras2} mounted on the 2.5-m telescope at the Physical Research Laboratory (PRL) Observatory, Gurushikhar, Mount Abu, Rajasthan, India. The instrument is equipped with a TRIUS PRO-814 CCD detector (model ICX814AL\footnote{\url{https://www.sxccd.com/product/trius-sx814//}}), which has $3388\times2712$ pixels, with a pixel size of $3.69\mu\text{m}\times3.69\mu\text{m}$. The field of view (FOV) of the system is $2.15'\times1.7'$, corresponding to a pixel scale of $38\,\text{mas/pixel}$. The observations were obtained using a Bessel-V filter. A total of approximately 2500 speckle frames were recorded, each with an exposure time of $50\,\text{ms}$, under favourable sky conditions with an average seeing of $\sim0.9''$. The data reduction was performed using a custom pipeline developed in \texttt{python}, following procedures similar to those described in \citet{2020ziegler} and \citet{2018tokovinin}. The analysis involved computing the power spectral density (PSD) function, constructing the autocorrelation function (ACF), and deriving the $5\sigma$ contrast limits. A detailed description of the reduction and analysis methodology is provided in \citet{toi6651}. The derived contrast limits for TOI-7154 are $\Delta V=5.89$ and $\Delta V=5.51$ at angular separations of $0.25''$ and $1.0''$, respectively. No stellar companions were detected. The top panel in Figure \ref{fig:speckle} presents the $5\sigma$ contrast curve along with the corresponding ACF for TOI-7154 obtained with this instrument.

\subsubsection{{{Speckle interferometry with SPeckle Polarimeter}}}
TOI-7154 was observed on March 13, 2025 with the {{SPeckle Polarimeter (SPP)} on the 2.5-m telescope at the Caucasian Observatory of Sternberg Astronomical Institute (SAI) of Lomonosov Moscow State University. A low-noise CMOS detector, Hamamatsu ORCA-quest \citep{Strakhov2023}, was used for the observations. The atmospheric dispersion compensator was active, which allowed us to use the wide $I_\mathrm{c}$ band, with an angular resolution of $0.083''$. Long exposure atmospheric seeing was $1.01''$ at the moment of observation. {{We calculated the ACF to estimate the detection limits depending on distance to the star. Then we calculated the standard deviation of the latter in annular regions having radii ranging from $0.05''$ to $1.7''$ and width of $0.06''$. We assumed that the potential companion would be detectable if its signal in ACF was larger than 5 standard deviations. The detection limit in ACF computed this way was converted into the magnitude difference, at distances $0.25''$ and $1.0''$ from the star, which amounted to $\Delta=3.5^m$ and $5.3^m$, respectively. For further details of this procedure see \citet{SPP2023}. The bottom panel in Figure~\ref{fig:speckle} presents the contrast curve and the corresponding ACF obtained using the SPP.}

\subsection{Reconnaissance Spectroscopy and RV Follow-up}

\subsubsection{PARAS-2 Observations}{\label{sec:paras2_obs}}
The RV follow-up of TOI-7154 was carried out using the PARAS-2 \citep{paras2_design,prl2.5m_and_paras2} spectrograph attached to the 2.5-m telescope at the PRL Mount Abu Observatory, Gurushikhar, Rajasthan, India. PARAS-2 is a fiber-fed high-resolution ($R\approx110000$) cross-dispersed echelle spectrograph kept under highly stable environmental conditions. The spectrograph works in the visible wavelength range of $3800\AA-6900\AA$. The spectrograph utilizes the Uranium-Argon (UAr) hollow cathode lamp for wavelength calibration and simultaneous referencing \citep{toi6651, uar}.  More details about the spectrograph can be found in \citet{paras2_design,prl2.5m_and_paras2}. We conducted a total of $20$ spectroscopic observations of TOI-7154, spanning $24$ days from March 17, 2025, to April 10, 2025, with an exposure time of $3600$ seconds for each observation. These spectra were extracted using the PARAS-2 pipeline, which is briefly explained later in the section. The RV errors are calculated using the techniques mentioned in \citet{Chaturvedi_2016, Chaturvedi_2018}. All the details of RVs, timestamps, and their corresponding errors are presented in Table~\ref{tab:rv_table}. The PARAS-2 data reduction pipeline, written in the \texttt{IDL} language, is an upgraded version of the PARAS pipeline \citep{Chakraborty2014}, which employs the \texttt{REDUCE} package \citep{Piskunov2002} for extracting cross-dispersed echelle spectra. Using calibration and science frames, the pipeline performs optimal spectral extraction through bias and flat corrections, order tracing via \texttt{REDUCE}’s clustering algorithm, and scattered-light estimation. Cosmic rays are removed using the Laplacian edge detection method \citep{Dokkum2001}, and spectra are extracted with a swath-based decomposition approach. Wavelength calibration is achieved using a Uranium line-list \citep{uar}, where a four-degree polynomial is fitted per order to derive the wavelength solution. The spectra are then cross-correlated with a G2-type mask following \cite{Baranne1996} and \cite{Pepe_ccf_2002}. More details on the PARAS-2 pipeline can be found in \citet{toi6651}.

\subsubsection{TRES Observations}{\label{sec:tres_obs}}

We also obtained 8 spectra of TOI-7154 between September 08, 2025, and October 30, 2025, using the Tillinghast Reflector Echelle Spectrograph (TRES; \citet{gaborthesistres}) on the $1.5\,\text{m}$ Tillinghast Reflector telescope on Mount Hopkins, Arizona, USA. TRES is a fiber-fed echelle spectrograph with a resolving power of R $\approx$ 44,000 that operates in the wavelength range of 390–910 nm. The observations were obtained in sequences of three stellar exposures surrounded by ThAr calibration exposures through the same fiber, to assist with the removal of cosmic rays. The total exposure times were in the range $1600\,\text{s}$ to $2520\,\text{s}$ and yielded S/N ratios per resolution element between 26 and 37. The spectra were extracted using the procedures outlined in \citet{tres2010}. Multi-order relative velocities were 
derived by cross-correlating each observed spectrum order by order against a master template constructed by combining all the TRES observations of TOI-7154. The estimated error for each observation was derived from the scatter of the RVs across the echelle orders for that observation. The RVs acquired with TRES spectra, along with their respective errors, are listed in Table~\ref{tab:rv_table}.

\section{Analysis and Results}\label{sec:analysis}

\subsection{Stellar parameters from spectral synthesis}\label{sec:tres_spc}

The two high Signal-to-Noise Ratio (SNR) spectra acquired on September 8, 2025, and September 30, 2025, with TRES were used for estimating the physical parameters of the stars, e.g., effective temperature $T_{\text{eff}}$, Metallicity $\text{[Fe/H]}$, surface gravity $\log g$, and projected stellar rotational velocity $v\sin{i_\star}$. The SNR per resolution element of these spectra was $34.7$ and $37.9$, respectively. These parameters were derived by applying the Stellar Parameter Classification (SPC) software \citep{spc2012, spc2021} to individual TRES spectra, and the final parameters were obtained by taking the weighted average of the parameters obtained from these two spectra. The SPC utilizes a library of synthetic spectra spanning the wavelength range of $5020\AA-5320\AA$, centered on the Mg-b triplet. With SPC, we found $T_{\text{eff}} = 5589.0\pm50.0\,\text{K}$, $\text{[Fe/H]}=0.14\pm0.08\,\text{dex}$, $\log g= 4.54\pm0.10$, and $v\sin{i_\star}=3.30\pm0.5\,\text{km/s}$.

\begin{table}[t!]
\caption{Basic parameters of TOI-7154 system}
\label{tab:star_table}
\centering
\begin{tabular}{llll}
\hline\hline
\noalign{\smallskip}
Parameter&TOI-7154&Ref.\\
\noalign{\smallskip}\hline
\noalign{\smallskip}
\multicolumn{3}{l}{\textbf{Identifiers:}}\\
TIC&377622130&(1)\\
\textit{Gaia}DR3& 4548696887058100352  &(2)\\
2MASS& J17381688+1512485&(3)\\
\noalign{\smallskip}\hline
\noalign{\smallskip}
\multicolumn{3}{l}{\textbf{Astrometry:}}\\
$\alpha_{J2000}$ & 17:38:16.88  & (2)\\
$\delta_{J2000}$ & +15:12:48.73  & (2)\\
$\mu_{\alpha}$ (mas yr$^{-1}$) &  -11.072 $\pm$ 0.020  & (2)\\
$\mu_{\delta}$ (mas yr$^{-1}$) & 9.558 $\pm$ 0.022 & (2)\\
$\varpi${\textdagger} (mas) & 3.346 $\pm$ 0.020 &  (2)\\
$d$ (pc) & $292.901\pm0.63$ &  (2)\\
\noalign{\smallskip}\hline
\noalign{\smallskip}
\multicolumn{3}{l}{\textbf{Photometry$^{\textdagger}$:}}\\
$T$   & 11.9445 $\pm$ 0.0066  & (1)\\
$G$   & 12.4247 $\pm$ 0.0003  & (2)\\
$G_{BP}$   & 12.8087 $\pm$ 0.0007  & (2)\\
$G_{RP}$   & 11.8792 $\pm$  0.0005  & (2)\\
$J$   & 11.261 $\pm$ 0.021  & (3)\\
$H$   & 10.942 $\pm$ 0.027 & (3)\\
$K_{S}$ & 10.858 $\pm$ 0.021  & (3)\\
$W1$  &  10.808 $\pm$ 0.023  & (5)\\
$W2$  &  10.855 $\pm$ 0.021 & (5)\\
$W3$  &  10.84 $\pm$ 0.118  & (5)\\
\noalign{\smallskip}\hline
\noalign{\smallskip}
\multicolumn{3}{l}{\textbf{Spectroscopic parameters (TRES SPC):}}\\
$T_{\rm eff}$ (K) &$5589\pm50$ & (6)\\
$\log{g}$ (cgs) &$4.54\pm0.10$ & (6)\\
$[{\rm Fe/H}]$ (dex) &$0.14\pm0.080$ & (6)\\
$v\sin{i_*}$ ($km \ s^{-1}$) & $3.30 \pm 0.50$  & (6)\\
\noalign{\smallskip}\hline
\noalign{\smallskip}
\multicolumn{3}{l}{\textbf{Derived stellar parameters$^{\textdaggerdbl}$:}}\\
$M_*$ (\msun) &$0.939^{+0.047}_{-0.043}$ & (6)\\
$R_*$ (\rsun) &$0.949^{+0.032}_{-0.030}$ & (6)\\
$L_*$ (\lsun) &$0.782^{+0.036}_{-0.047}$ & (6)\\
$\rho_*$ (cgs) &$1.55^{+0.18}_{-0.16}$ & (6)\\
$T_{\rm eff}$ (K) &$5564^{+100}_{-110}$ & (6)\\
$\log{g}$ (cgs) &$4.456^{+0.036}_{-0.036}$ & (6)\\
$[{\rm Fe/H}]$ (dex) &$0.154^{+0.077}_{-0.075}$ & (6)\\
$Age$ (Gyr) &$7.2^{+3.9}_{-3.6}$ & (6)\\
$A_V$ (mag) &$0.173^{+0.052}_{-0.085}$ &  (6)\\
\noalign{\smallskip}
\hline
\noalign{\smallskip}
\end{tabular}\\
\footnotesize{\textbf{Note:} Parameters for TOI-7154 are obtained from the global modeling (see Section \ref{sec:global_modeling} for details). \textbf{References:} (1) \citet{stassun2018}, 
(2) \citet{gaia2023}, 
(3) \citet{cutri2003}, 
(4) \citet{hartman2020}, 
(5) \citet{cutri2021}, 
(6) This Work, and 
(7) \citet{paegert2021}.
}
\end{table}

\subsection{Rotation period of the star}\label{sec:star_rotation}
To determine the rotation period of the host star, we searched for long-term ground-based photometric observations from surveys such as SuperWASP \citep{superwasp}. However, no archival data was available for TOI-7154. Since the star was observed in multiple TESS sectors spanning a baseline of approximately 1460 days, we utilized these datasets to search for periodic signals distinct from the orbital period of the companion. We used the detrended flux (\texttt{DET} or \texttt{KSPSAP}) from several sectors. {The light curves were analyzed without additional long-term trend removal, after excluding contaminated portions as described in Section~\ref{sec:tess_obs}.} After masking the transits, we applied a Generalized Lomb-Scargle (GLS) periodogram to the combined dataset but did not detect any significant periodicity. The presence of substantial data gaps, caused by contamination from scattered sunlight, likely limited the sensitivity of the GLS algorithm to long-term signals. Therefore, we estimated the stellar rotation period using the projected rotational velocity derived from the SPC analysis (see Section \ref{sec:tres_spc}, $v\sin{i_\star}=3.30\pm0.5\,\text{km/s}$) and the stellar radius ($R_\star=0.949^{+0.032}_{-0.030}\,R_\odot$, see Section \ref{sec:global_modeling}). Assuming the star is viewed equator-on, this results in a rotation period of $13.5\pm2.0\,\text{d}$, which gives us the maximum limit of the period to be $\approx15.5\,\text{d}$.


\subsection{Periodogram analysis}
\begin{figure}[t!]
    \centering
    \includegraphics[width=\columnwidth]{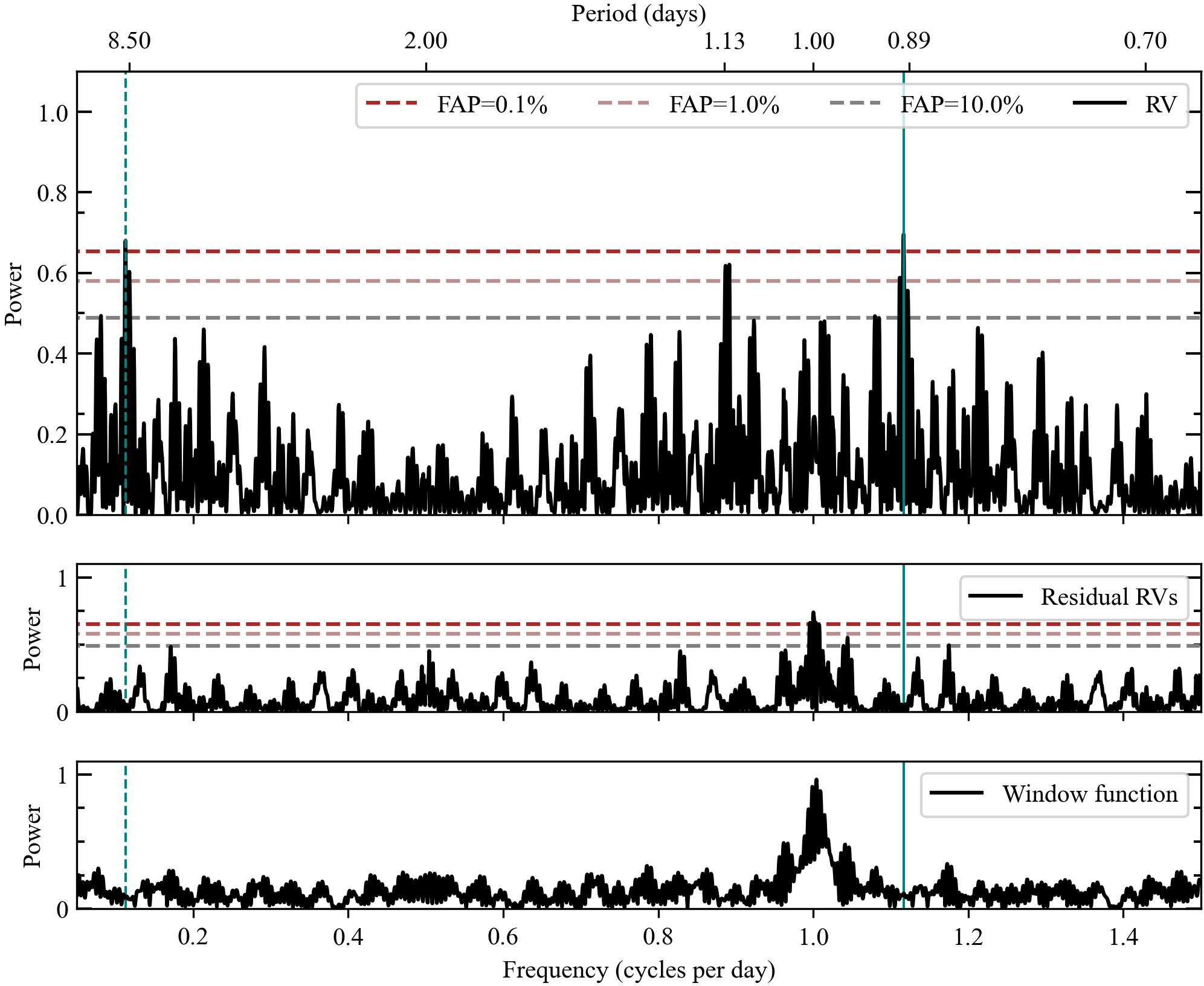}
    \caption{{The GLS periodogram for the combined PARAS-2 and TRES RVs, the residuals RVs, and the spectral window function of TOI-7154 are shown in panels 1–3 (top to bottom), respectively. The periodogram shows prominent peaks at $\sim0.89$~days (teal solid line) and $\sim8.5$~days (teal dashed line). The $\sim8.5$~day periodicity is consistent with the TESS transit period ($P \sim 8.86$~days) and represents the true orbital signal, while the $\sim0.89$~day peak is an alias. After removing the 8.5-day signal, no significant periodicities are found. The FAP levels corresponding to 0.1\%, 1\%, and 10\% are shown as horizontal grey, pink, and red dashed lines, respectively.}}
    \label{fig:periodogram}
\end{figure}

{We computed the GLS periodogram for the combined PARAS-2 and TRES RV data, shown in the upper panel of Figure~\ref{fig:periodogram}. The theoretical 0.1\%, 1\%, and 10\% false-alarm probability (FAP) levels are indicated by horizontal dotted lines. We adopted the 0.1\% FAP level as the threshold for identifying significant periodicities. The periodogram shows two prominent peaks above this threshold: a dominant peak at $\sim0.89$~days and a secondary peak at $\sim8.5$~days. The latter is consistent with the orbital period derived from the TESS light curve ($P \sim 8.86$~days), indicating that it corresponds to the true physical signal. The $\sim0.89$~day peak and a weaker peak near $\sim1.13$~days correspond to aliases of the orbital frequency, $f_{\rm orb}$ . Specifically, the $\sim0.89$~day and $\sim1.13$~day signals correspond to the $+f_{\rm orb}$ and $-f_{\rm orb}$ aliases, respectively. To verify this interpretation, we subtracted the $\sim8.5$~day signal using a best-fit sinusoidal model and computed the periodogram of the residuals (middle panel of Figure~\ref{fig:periodogram}). In the residual periodogram, all significant peaks, including those at $\sim0.89$ and $\sim1.13$~days disappear, confirming that they originated from the same underlying periodicity. The spectral window function is also shown in the lower panel of Figure~\ref{fig:periodogram} for the reference.}

\subsection{Global modeling}\label{sec:global_modeling}
We characterized the TOI-7154 system using the \texttt{EXOFASTv2} package\footnote{\url{https://github.com/jdeast/EXOFASTv2.git}} \citep{eastman2019}, a modeling and fitting suite written in the \texttt{IDL} language that simultaneously fits the RV and transit data. It also derives stellar properties by performing spectral energy distribution (SED) fitting \citep{stassun2016} in combination with MIST stellar evolutionary tracks \citep{choi2016}. The code utilizes a Markov Chain Monte Carlo (MCMC) framework, with convergence assessed through two criteria -- a Gelman–Rubin statistic below 1.01 and more than 1000 independent samples \citep{ford2008}.

\subsubsection{Stellar parameters}
Within the \texttt{EXOFASTv2} framework, the host star was modeled by a global SED fitting \citep{stassun2016, dtr2016} based on the Kurucz stellar atmosphere grids \citep{kurucz1979} with MESA Isochrones and Stellar Tracks (MIST; \citet{choi2016, dtr2016}). Combined with space-based precise transit photometry, this approach provides more accurate estimates of the stellar mass, radius, surface gravity, and age \citep{torres2008, eastman2023}. The TESS normalized light curve of TOI-7154 is used to model the host star. For SED fitting, we used the Broadband photometric magnitudes from Gaia (G, GBP, GRP), 2MASS (J, H, K), and ALLWISE (W1, W2, W3) (see Table~\ref{tab:star_table}). {A minimum uncertainty floors were applied to every photometric measurements to avoid over-weighting individual bands, and a global scaling factor ($\sigma_{SED}$) was fitted to ensure appropriate weighting of the SED within the global modelling. Systematic floors in the stellar parameters, $T_{\mathrm{eff}}$ and $F_{\mathrm{bol}}$, were incorporated following \citet{Tayar2022}, ensuring that the SED does not over-constrain the stellar parameters within the global fit.} To better constrain the model parameters, Gaussian priors were applied to the metallicity $\text{[Fe/H]}$, based on spectral synthesis results (see Section~\ref{sec:tres_spc}), and to the Gaia DR3 parallax \citep{gdr3} after applying the systematic corrections of \citet{lindegren2021}. No prior constraints were enforced on $\log g$ or $T_{\text{eff}}$, although the median values from the spectroscopic analysis (see Section~\ref{sec:tres_spc}) were applied as starting points. In addition, a uniform prior was imposed on the V-band extinction using the dust maps of \citet{schlafly2011} at the position of TOI-7154.

Figure~\ref{fig:mist} shows the $T_{\text{eff}}-\log g$ relation corresponding to the best-fit model, along with the best-fit MIST evolutionary track for TOI-7154, plotted as a solid curve. The broadband SED fit to the observed photometric fluxes is presented in Figure~\ref{fig:sed}. The stellar and planetary parameters inferred from \texttt{EXOFASTv2} -- reported as medians with 68\% confidence intervals (1$\sigma$ uncertainties) -- are listed in Tables~\ref{tab:star_table} \&~\ref{tab:planet_table}, respectively. For the best-fit solution, the most probable host-star parameters are -- $M_\star= 0.939^{+0.047}_{-0.046}\,M_\odot$, $R_\star=0.949^{+0.032}_{-0.030}\,R_\odot$, $T_{\text{eff}}=5564^{+100}_{-110}\,\text{K}$, $\log g=4.456\pm{0.036}$, and $\text{[Fe/H]}= 0.154^{+0.077}_{-0.075}\,\text{dex}$, at an age of $7.2^{+3.9}_{-3.6}\,\text{Gyr}$. These results are consistent with the values obtained from independent spectroscopic analysis (see Section~\ref{sec:tres_spc}).

\begin{figure}[t!]
    \centering
    \includegraphics[width=\columnwidth]{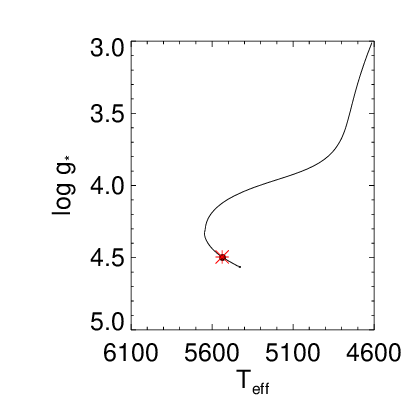}
    \caption{MIST evolutionary track for TOI-7154 shown as a solid black line. The black point indicates the $T_\mathrm{eff}$ and $\log{g}$, while the red asterisk denotes the current age of TOI-7154.}
    \label{fig:mist}
\end{figure}

\begin{figure}[t!]
    \centering
    \includegraphics[width=\columnwidth]{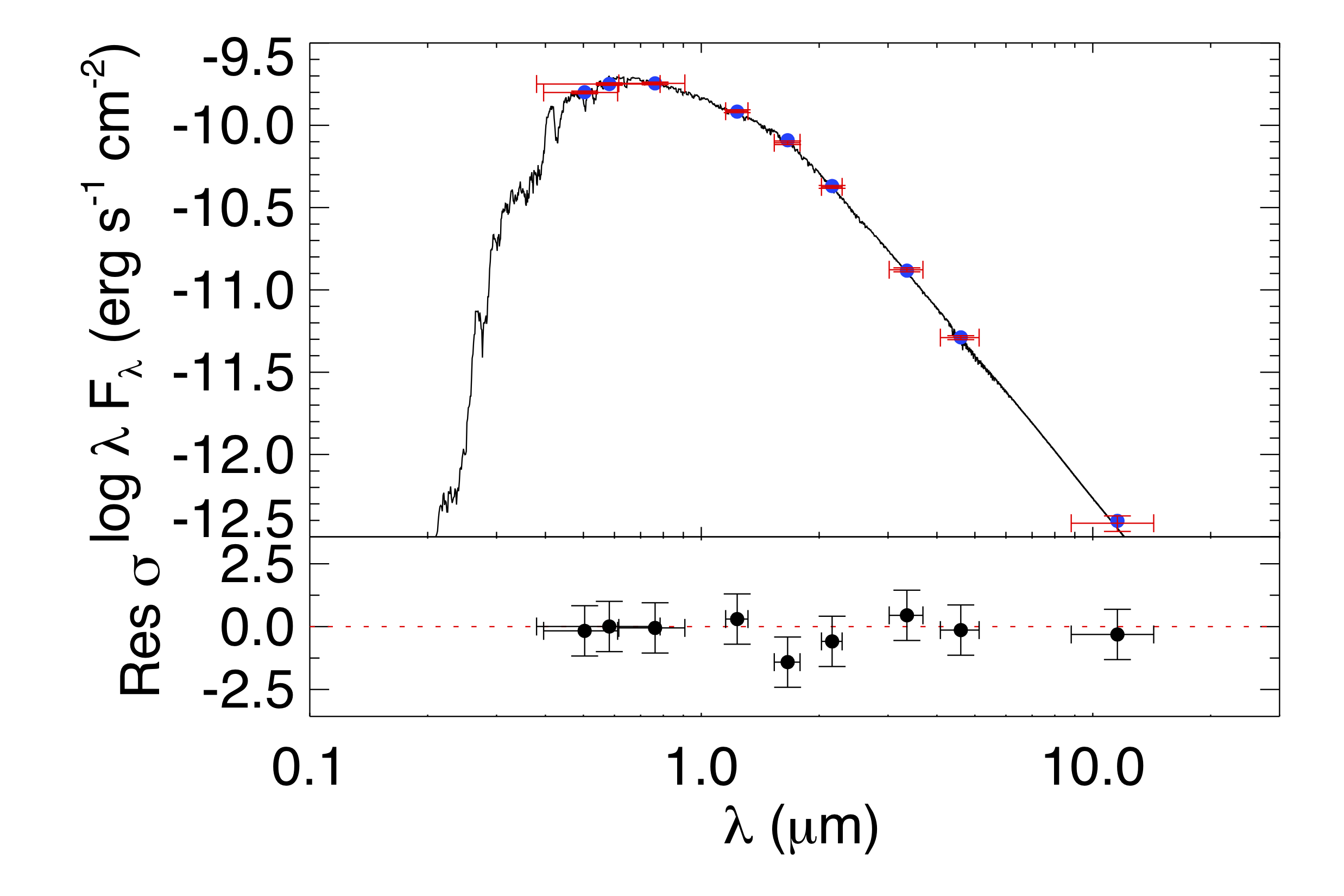}
    \caption{Spectral energy distribution of TOI-7154. The observed photometric measurements are represented by the red symbols, while the horizontal bars are indicative of the effective width of the respective passband. The model fluxes are represented by blue points. The residuals are shown in the lower panel.}
    \label{fig:sed}
\end{figure}

\subsubsection{Planetary parameters}
The joint analysis of the TESS light curves and PARAS-2 and TRES radial velocities was carried out by treating all orbital parameters -- impact parameter ($b$), inclination ($i$), semi-major axis ($a$), planetary radius ($R_p$), RV semi-amplitude ($K$), eccentricity ($e$), and argument of periastron ($\omega$) -- as free variables. Only starting values of the orbital period ($P$) and mid-transit time ($T_c$) were provided from the TESS QLP pipeline. Transit modeling followed the analytic framework of \citet{mandel2002}, while the RVs were described with a non-circular Keplerian solution. {Independent RV zero-point offsets were also included for TRES and PARAS-2 spectrograph and fitted simultaneously with the orbital parameters.}
A quadratic limb-darkening law was adopted for the TESS passband, with coefficients $(u_1, u_2)$ interpolated from the tables of \citet{claret2011} and \citet{claret2017}. For TOI-7154, TESS observations span four sectors with different integration times. To account for long-cadence smearing, the transit model for the $1800\,\text{s}$ exposures in Sector 79 was resampled over 10 sub-steps, and for the $600\,\text{s}$ exposures in Sectors 52 and 53, it was resampled over 3 sub-steps. The combined modeling yields a sub-stellar companion (brown dwarf) with a radius of $R_P=0.827^{+0.040}_{-0.037}\,R_{\mathrm{J}}$ and a mass of $M_P=71.7^{+2.4}_{-2.2}\, M_{\mathrm{J}}$, corresponding to an RV semi-amplitude of $K=7232^{+35}_{-33}\,\text{m/s}$ and an orbital eccentricity of $e=0.2482\pm0.0024$. The median values and 68\% confidence intervals (1$\sigma$) for the best-fit solution planetary parameters are reported in Table~\ref{tab:planet_table}. The phase-folded TESS light curves and PARAS-2 and TRES RV data, along with their best-fit models, are displayed in Figures~\ref{fig:tesslc} and \ref{fig:rv}, respectively.
\begin{figure*}[ht!]
    \centering
    \includegraphics[width=0.49\textwidth]{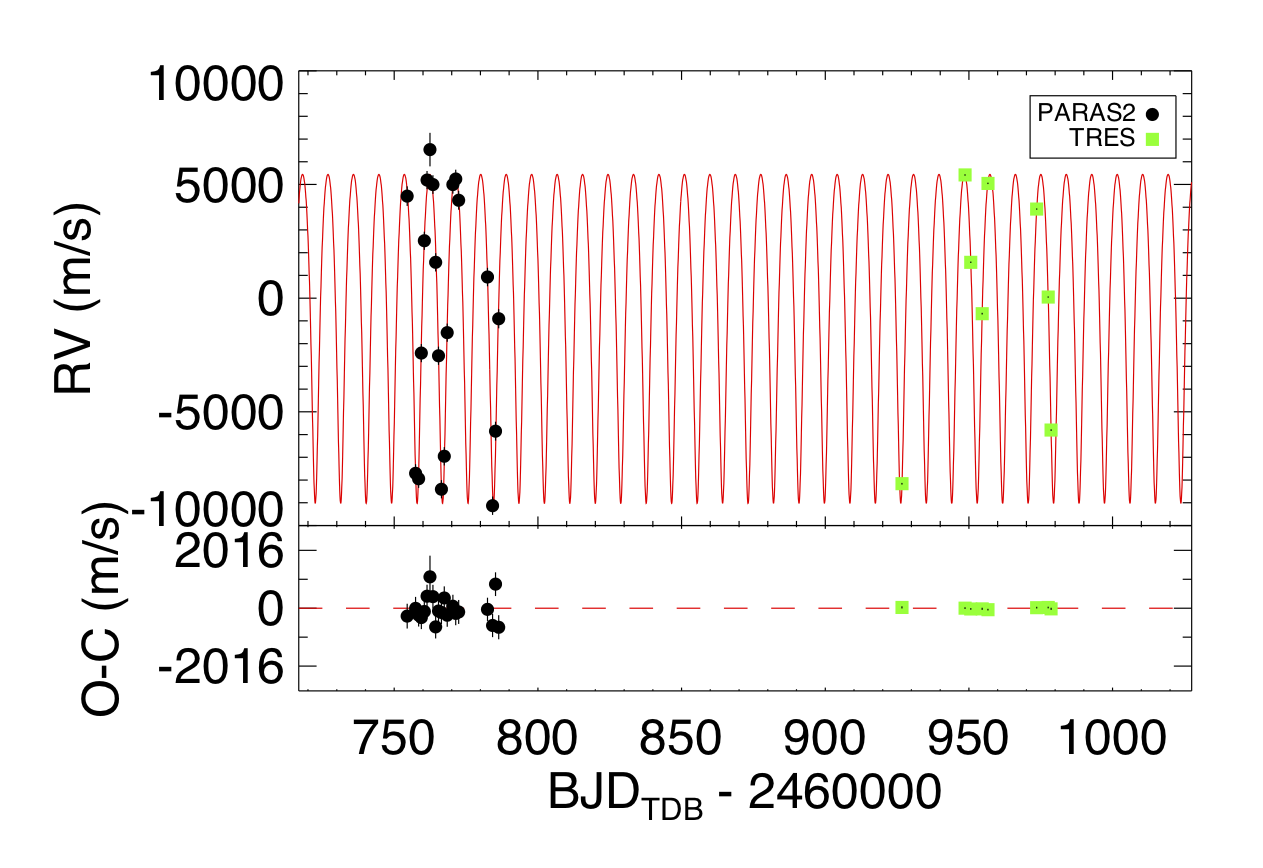}
    \includegraphics[width=0.49\textwidth]{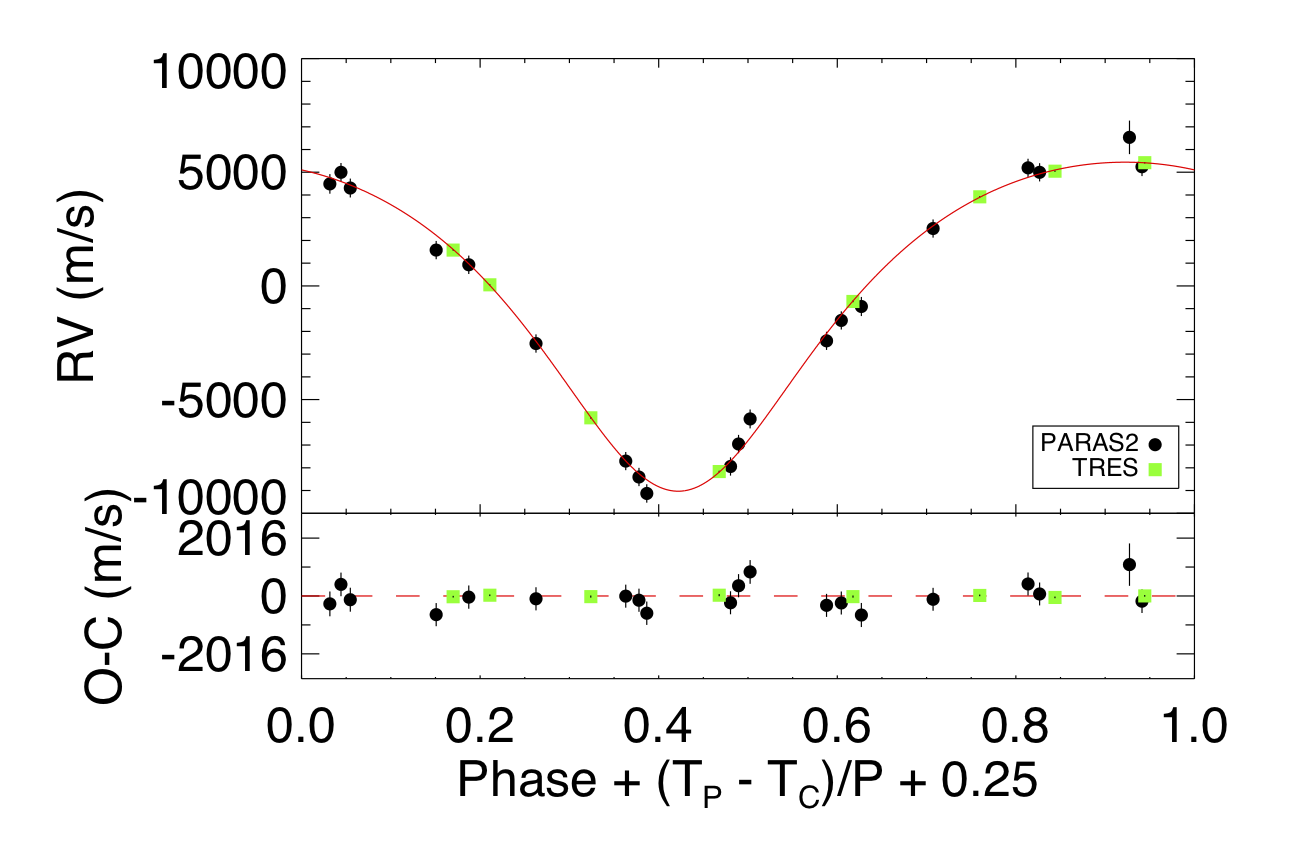}
    \caption{RV measurements of TOI-7154 from 20 spectroscopic observations with PARAS-2 (black points) and 8 spectroscopic observations with TRES (green squares) are shown as a function of time (left panel). The same RV data are shown as a function of orbital phase (right panel). In both panels, the best-fit RV model obtained using EXOFASTv2 is shown by the red curve, while the residuals between the model and the observations are displayed in the corresponding lower panels.}
    \label{fig:rv}
\end{figure*}

\subsection{System Age}
Since the uncertainties associated with age estimation from MIST isochrones are {large} ($\ge 50 \%$), we employed several other techniques to constrain the system's age. 

\subsubsection{Age estimation using Gyrochronology}
We use the derived rotation period $P_{rot}$ and Gaia $B_P$ and $R_P$ colors (Table~\ref{tab:star_table}), following the empirical gyrochronology relations of \citet{stardate} implemented via the {\tt stardate} package\footnote{\url{https://github.com/RuthAngus/stardate.git}}. This yielded a gyrochronological age of $1.41\pm0.34~\mathrm{Gyr}$. The relatively young age inferred from rotation contrasts with expectations based on the star’s effective temperature and surface gravity. One possible explanation is that the empirical relations of \citet{stardate} are calibrated primarily for single, isolated stars. Given the presence of a close-in brown dwarf companion of $\approx 72\ M_{\mathrm{J}}$ in an 8.86-day orbit, tidal or orbital interactions may have spun up the star, resulting in an anomalously short rotation period and, consequently, a younger gyrochronological age \citep{gyroage_planets, EB_rotation2017}.

\subsubsection{Age estimation using Galactic kinematics}
{To derive the system's age using galactic kinematics, we followed the method described in \citet{Bensby+2014}}, wherein we utilize the radial velocity estimates obtained using the PARAS-2 data along with the position, proper motion, and distance estimates from Gaia DR3 \citep{gdr3}. We calculate the Galactic space velocity components -- $U$, $V$ and $W$, for TOI-7154 using the \verb|gal_uvw| function of the \verb|Pyastronomy.pyasl| python module. We also applied corrections relative to the Local Standard of Rest (LSR) by taking the Sun's Galactic space velocity components $(U_\odot, V_\odot, W_\odot)=(11.1, 12.24, 7.25)\;\text{km/s}$ from \citet{Bensby+2014}, where we follow the convention of $U$ being positive towards the Galactic center, $V$ being positive in the direction of Galactic rotation and $W$ being positive towards the North Galactic Pole. Using this approach, we obtained the Galactic spatial velocity of TOI-7154 as $U_{\text{LSR}}=-18.102\pm0.36\;\text{km/s}$, $V_{\text{LSR}}=-3.407\pm0.29\;\text{km/s}$ and $W_{\text{LSR}}=14.927\pm0.19\;\text{km/s}$.

Furthermore, we borrowed the nomenclature scheme and built up on the procedure in \citep{Bensby+2014} using the values of asymmetric drift velocities -- $U_\text{asym}$ and $V_\text{asym}$, velocity dispersions -- $\sigma_U$, $\sigma_V$ and $\sigma_W$, and the normalisation fraction $X$ for different populations in the Solar neighbourhood and the LSR, to calculate the Galactic space velocity distribution which we then multiply with the observed fraction $X$ to get the probability that TOI-7154 belongs to a specific Galactic population. The relative thick--thin disk $(P_{TD}/P_D)$, thick disk--halo $(P_{TD}/P_H)$ and thin disk--halo $(P_D/P_H)$ probabilities are then calculated by dividing the respective individual probability estimates. The obtained relative probabilities are : $P_{TD}/P_D=0.0131$, $P_{TD}/P_H=7569.4904$ and $P_D/P_H=579513.9929$. Based on the discussion in \citet{Henderson+2024} using the interpretations of the relative probability estimates in \citet{Bensby+2014} and the average age of different Galactic populations as discussed in \citet{Prantzos+2023}, we can infer that TOI-7154 is a part of the thin disk population having an average age of $\sim4-5\;\text{Gyr}$. This age estimate, however lies at the lower edge of the estimate obtained by the MIST isochrone fitting.


\section{Discussion}\label{sec:discussion}

\subsection{The Mass–Radius Relation of Brown Dwarfs: Position of TOI-7154b}
\begin{figure}[ht!]
\centering
\includegraphics[width=1.1\columnwidth]{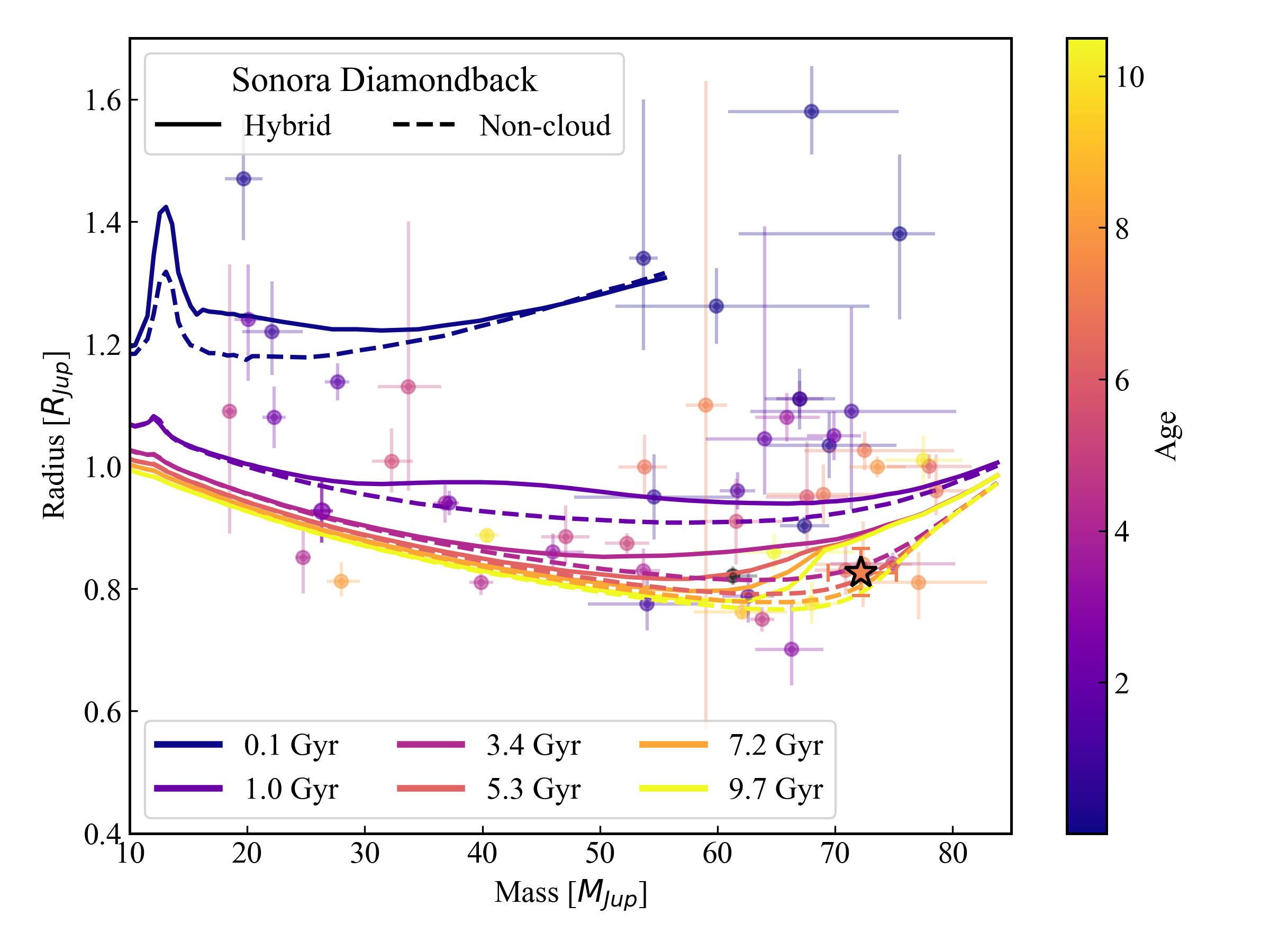}
\caption{{Mass--radius diagram of the known transiting BDs, including TOI-7154b (represented by star), compared with the \texttt{Sonora Diamondback} evolutionary models (SM24, \citet{SM24}). The sample of transiting brown dwarfs compiled from the \texttt{TEPCAT} database (see \citet{tepcat} and references therein) and represented by points colored according to their age. These SM24 models are at solar metallicity ($Z=0.0$), and the hybrid (solid) and the non-cloud (dashed) model tracks are plotted at ages of 0.1, 1.0, 3.4, 5.3, 7.2, and 9.7~Gyr. Hybrid models include the effects of clouds on the atmospheric structure and cooling of the BDs, whereas non-cloud models assume that condensates have settled below the photosphere. TOI-7154b lies closest to the non-cloud models, indicating that its evolution is best described by cloud-free models. Its mass ($71.7~M_{\mathrm{J}}$) and radius ($0.827~R_{\mathrm{J}}$) place it near the minimum of the brown dwarf mass-radius relation, consistent with strong electron-degeneracy support and an old age.}}
\label{fig:massradius}
\end{figure}

{We show the position of TOI-7154b in the mass–radius plane of BDs in Figure~\ref{fig:massradius}, alongside known transiting BDs with masses between 13~$M_{\mathrm{J}}$ and 80~$M_{\mathrm{J}}$. The comparison sample is drawn from the \texttt{TEPCAT} database \citep{tepcat}, including the recently characterized BD TOI-6884b (Khandelwal et al.\ under review). We overplot the \texttt{Sonora Diamondback} evolutionary model tracks (SM24, \citet{marley2021, SM24}) for a range of ages. The observed distribution reproduces the well--known BD mass–radius relation, in which radii decrease with increasing mass up to $\sim60$–$70\,M_{\mathrm{J}}$. This behavior reflects the increasing dominance of electron degeneracy pressure over thermal pressure in the interiors of these objects \citep{chabrier2000, burrows2001, chabrier2014}. As the mass increases, degeneracy strengthens, leading to further contraction. At the highest masses, approaching the hydrogen--burning limit ($\sim75\, M_{\mathrm{J}}$), this trend begins to flatten and eventually reverses as stable hydrogen fusion becomes possible. Throughout its life, the mass of a BD remains nearly constant over time, but the radius evolves significantly as it cools and contracts. Younger BDs can retain radii close to that of Jupiter, whereas older objects ($\gtrsim$1–5~Gyr) contract to radii below $\sim1\,R_{\mathrm{J}}$ \citep{baraffe2003, Burrows2011, Chabrier2023}. }

{TOI-7154b, with a mass of $71.7^{+2.4}_{-2.2}~M_{\mathrm{J}}$, radius of $0.827^{+0.040}_{-0.037}~R_{\mathrm{J}}$, and an estimated age of $7.2^{+3.9}_{-3.6}$~Gyr, is consistent with these evolutionary models. Its position relative to the isochrones implies an age of $\gtrsim$3~Gyr, in agreement with the inferred stellar age with isochrones and Galactic Kinematics. Despite its proximity to the host star, TOI-7154b does not exhibit evidence for radius inflation. The incident stellar flux on the BD is $\sim1.4 \times 10^8$~erg~s$^{-1}$~cm$^{-2}$ ($\log_{10}(F) \approx 8.15$), which lies below the threshold ($\log_{10}(F) \gtrsim 9$ or $T_{\mathrm{eq}} \gtrsim 1450$~K) required for significant irradiation-driven inflation \citep{Mukherjee2026}. This indicates that irradiation effects are not expected to play a major role in inflating its radius. The corresponding high mean density of $157^{+25}_{-22}~g~cm^{-3}$ further indicates a compact, high-gravity BD that has undergone substantial contraction as expected for an old BD. Its properties are well reproduced by the SM24 non-cloud evolutionary models, and its location along these tracks corresponds to an effective temperature of $T_{\mathrm{eff}} \gtrsim 1500$~K, placing it in the T-dwarf regime.}

{To place TOI-7154b in the context of the broader BD population, we compared it with some objects of similar mass ($\sim70$–$80\,M_{\mathrm{J}}$) and age ($\gtrsim$3~Gyr), such as  EPIC~201702477b \citep{Bayliss+2017}, TOI-5575b \citep{Gan+2025}, TOI-2533b \citep{dos-Santos+2024}, TOI-148b \citep{Grieves2021} and Kepler-503b \citep{Canas+2018}. While most of these systems exhibit radii in the range $\sim0.83$–$0.95\, R_{\mathrm{J}}$, consistent with evolutionary models, a few outliers with inflated radii exist. For example, TOI-2521b ($R = 1.01 \pm 0.04\, R_{\mathrm{J}}$; \citealt{Lin2023}) and TOI-2490b ($R = 0.999 \pm 0.017\, R_{\mathrm{J}}$; \citealt{Henderson2024}), among others, appear slightly inflated despite their mature ages. The inflation in TOI-2521b is likely driven by the higher incident flux ($\log_{10}(F) \approx 9.0$, \citet{Lin2023}), consistent with a recent study by \citet{Mukherjee2026}. In contrast, the inflation in TOI-2490b may be linked with its highly eccentric orbit ($e \sim 0.8$), which results in significant variations in incident flux as it goes from periastron to apoastron. This could contribute to delayed contraction through enhanced atmospheric heating, although the exact origin of the inflation remains uncertain \citep{Henderson2024}. Such behavior is not evident in TOI-7154b and the aforementioned older BDs, although the current sample size is very small. Furthermore, the observed scatter in the mass-radius diagram for known transiting BDs could arise from a combination of factors, including differences in age, metallicity, stellar irradiation, and observational uncertainties. In addition, the lack of precise age estimates for many systems leads to discrepancies between the systems' ages and evolutionary models, further contributing to this apparent dispersion. In the context, TOI-7154b serves as an important empirical benchmark for testing evolutionary models and constraining population-level trends for the objects near the hydrogen-burning mass limit.}

\subsection{Tidal Evolution}\label{sec:tidal1}
\begin{table}[t!]
\caption{Tidal circularisation timescales for TOI-7154 system estimated using equations (\ref{eq:tidal3}-\ref{eq:tidal5}).}
\label{tab:tidal1}
\centering
\begin{tabular}{llccc}
\hline \hline 
\noalign{\smallskip}
$Q_\star'$ & $Q_p'$ & $\tau_{\text{circ}}^{\star}$ (Gyr) & $\tau_{\text{circ}}^{\text{p}}$ (Gyr) & $\tau_{\text{circ}}^{\text{eff}}$ (Gyr) \\ 
\noalign{\smallskip} \hline
\noalign{\smallskip}
$10^5$ & $10^{4.5}$ & $1.35$ & $293.71$ & $1.34$ \\
 & $10^{5}$ & $1.35$ & $928.80$ & $1.35$ \\
 & $10^{5.5}$ & $1.35$ & $2937.11$ & $1.35$ \\
 & $10^{6}$ & $1.35$ & $9287.97$ & $1.35$ \\
 & $10^{6.5}$ & $1.35$ & $29371.13$ & $1.35$ \\
\noalign{\smallskip} \hline
\noalign{\smallskip}
$10^{5.5}$ & $10^{4.5}$ & $4.27$ & $293.71$ & $4.20$ \\
 & $10^{5}$ & $4.27$ & $928.80$ & $4.25$ \\
 & $10^{5.5}$ & $4.27$ & $2937.11$ & $4.26$ \\
 & $10^{6}$ & $4.27$ & $9287.97$ & $4.26$ \\
 & $10^{6.5}$ & $4.27$ & $29371.13$ & $4.27$ \\
\noalign{\smallskip} \hline
\noalign{\smallskip}
$10^{6}$ & $10^{4.5}$ & $13.49$ & $293.71$ & $12.90$ \\
 & $10^{5}$ & $13.49$ & $928.80$ & $13.30$ \\
 & $10^{5.5}$ & $13.49$ & $2937.11$ & $13.43$ \\
 & $10^{6}$ & $13.49$ & $9287.97$ & $13.47$ \\
 & $10^{6.5}$ & $13.49$ & $29371.13$ & $13.48$ \\
\noalign{\smallskip} \hline
\noalign{\smallskip}
$10^{7}$ & $10^{4.5}$ & $134.9$ & $293.71$ & $92.44$ \\
 & $10^{5}$ & $134.9$ & $928.80$ & $117.79$ \\
 & $10^{5.5}$ & $134.9$ & $2937.11$ & $128.97$ \\
 & $10^{6}$ & $134.9$ & $9287.97$ & $132.96$ \\
 & $10^{6.5}$ & $134.9$ & $29371.13$ & $134.28$ \\
\noalign{\smallskip} \hline
\end{tabular}
\end{table}
We adopt the classical equations governing tidal evolution from \citet{Jackson+2009}, which take into account the inter-dependent evolution of orbital separation, $a$, and eccentricity, $e$, in the form of a system of coupled first-order ordinary differential equations, given by
\begin{equation}\label{eq:tidal1}
    \frac{1}{a}\frac{\mathrm da}{\mathrm dt}=-\left[\frac{63}{2}\left(\frac{R_p^5\sqrt{GM_\star^3}}{Q_p'M_p}\right)e^2+
    \frac{9}{2}\sqrt{\frac{G}{M_\star}}\frac{R_\star^5M_p}{Q_\star'}\left(1+\frac{57}{4}e^2\right)\right]a^{-\frac{13}{2}}
\end{equation}
\begin{equation}\label{eq:tidal2}
    \frac{1}{e}\frac{\mathrm de}{\mathrm dt}=-\left[\frac{63}{4}\sqrt{GM_\star^3}\left(\frac{R_p^5}{Q_p'M_p}\right)+ \\
    \frac{225}{16}\sqrt{\frac{G}{M_\star}}\left(\frac{R_\star^5M_p}{Q_\star'}\right)\right]a^{-\frac{13}{2}} 
\end{equation}
where $G$ is the universal gravitational constant, $R_{\star, p}$ denote the radii, $M_{\star, p}$ denote the masses and $Q_{\star,p}$ represent the modified tidal dissipation parameters for the host star and the planet, respectively. The various underlying assumptions for the validity of these equations are detailed in \citep{Jackson+2008, Jackson+2009}. The major assumptions that govern their implementation in the current work are :
\begin{enumerate}
    \item The rotation period of the host star should be at least two-third the orbital period of the planetary body -- the failure of this assumption changes the constant terms and behavior of tidal dissipation factors
    \item Synchronous (or nearly-synchronous) orbit of the planet around the host star
    \item The orbital separation should be smaller than $0.2\;\text{AU}$ -- regime of validity of the derived results \citep{Jackson+2008} 
    \item {The orbital eccentricity should be small -- in the regime where this assumption no longer holds, orbital harmonics should be introduced \citep{Goldreich1963}. In the absence of accounting such interactions, the estimates remain conservative.}
\end{enumerate}
In the present scenario, all the aforementioned assumptions hold good. {For the case of stellar/sub-stellar interactions as in the present work, the eccentricity of 0.2 lies in the moderately-low regime, where the orbital harmonics can be safely neglected \citep{KoenigsbergerEstrella-Trujillo2024}}. 

The choice of $Q'$ parameters, however, requires a separate discussion altogether. The $Q_\star'$ terms reflect the effects of stellar tides (tides on the host star due to the planet) while the $Q_p'$ terms reflect the effects of planetary tides (tides on the planet due to the host star). The choice of $Q_p'$ varies from $10^{4.5}$ \citep{Yoder+1981} to $10^6$ \citep{Bodenheimer+2003} with the effective value of $5\times10^6$ for close-in extrasolar planets \citep{Ogilvie+2004}. For gaseous planets, the tidal quality factor has been estimated to be in the range $10^6<Q_p'<10^8$ \citep{Jackson+2008, Hansen2010, Hansen2012, Quinn2014, Fellay2023}. For $Q_\star'$, the adopted values range from $10^5$ to as high as $10^9$ \citep{Lin+1996, Carone+2007, penev2012, barker2010, essick2016, Hamer2019, Patel2022}. {Additionally, for a system similar to TOI-7154 -- CWW 89A, a comparatively less massive brown dwarf ($M\sim36\,M_{\mathrm J}$) orbiting a Sun-like star, \citet{Beatty2018} used orbital and rotational constraints to infer $Q_p'>10^{4.5}$ and $Q_\star'>10^9$. It is therefore, evident that the choice of $Q'$ varies from system to system and strongly depends on the body's internal structure. An in-depth analysis is needed to adopt the right estimate for the system under consideration.}

In most of the short-period exoplanet systems, however, the effect of stellar tides in tidal evolution is insignificant as dissipation within the planet is usually expected to dominate over dissipation in the star. In such a simplistic scenario, we can neglect the circularisation effect of stellar tides and furthermore, assume the orbital separation to be constant \citep{Trilling2000, Bodenheimer+2003, Henderson+2024} and estimate a \textit{circularisation timescale} for the system as
\begin{eqnarray}
    \frac{1}{\tau_{\text{circ}}^\star}&=&\frac{225}{16}\sqrt{\frac{G}{M_\star}}\left(\frac{R_\star^5M_p}{Q_\star'}\right)a^{-\frac{13}{2}}\label{eq:tidal3} \\
    \frac{1}{\tau_{\text{circ}}^\text{p}}&=&\frac{63}{4}\sqrt{GM_\star^3}\left(\frac{R_p^5}{Q_p'M_p}\right)a^{-\frac{13}{2}}\label{eq:tidal4} \\
    \tau_{\text{circ}}^{\text{eff}}&=&\left(\frac{1}{\tau_{\text{circ}}^\star}+\frac{1}{\tau_{\text{circ}}^p}\right)^{-1}\approx\tau_{\text{circ}}^p\,\, for\,\,Q'_\star\gg Q'_p\label{eq:tidal5}
\end{eqnarray}
{\citet{Henderson+2024} used equations (\ref{eq:tidal3}-\ref{eq:tidal5}) to comment on the tidal evolution of TOI-2490b, which has a mass of about $74\,M_{\mathrm{J}}$ and an orbital eccentricity of about $0.78$. Such an analysis is not ideal for systems with massive BDs since the effect of stellar tides becomes significant. Additionally, the high orbital eccentricity requires additional orbital harmonics to be considered while modeling the tidal evolution of such a system. \citet{Jackson+2008} discusses the serious impact of neglecting stellar tides on the estimate of circularisation timescale (see Figure 2 therein). For comparison, we performed analysis on similar grounds as done in \citet{Henderson+2024} and the obtained estimates for $\tau_{\text{circ}}^{\text{eff}}$ are summarised in Table \ref{tab:tidal1}.}

We also performed a complete tidal evolution of the system by numerically integrating the coupled differential equations (\ref{eq:tidal1}-\ref{eq:tidal2}) incorporating the effects of both stellar and planetary tides. We have taken the mass and radius estimates along with the initial orbital separation and eccentricity values from the results of global modeling with \texttt{EXOFASTv2} (shown in Tables \ref{tab:star_table} and \ref{tab:planet_table}). We have taken three representative values of $Q_\star'$ as $10^{5.5}$, $10^{6}$ and $10^{6.5}$ and evolved the system for varying $Q_p'$ values, both in the forward and backward direction, to demonstrate the influence of stellar tides in the evolution of the system. We have also considered the limiting case of $Q_p'\to\infty$ for theoretical understanding. {The above choices of $Q'_\star$ are adopted to make sure the impact of stellar tides is significant, as one may expect from a massive close-in BD as TOI-7154b.}
\begin{figure*}[t!]
    \centering
    \includegraphics[width=\linewidth]{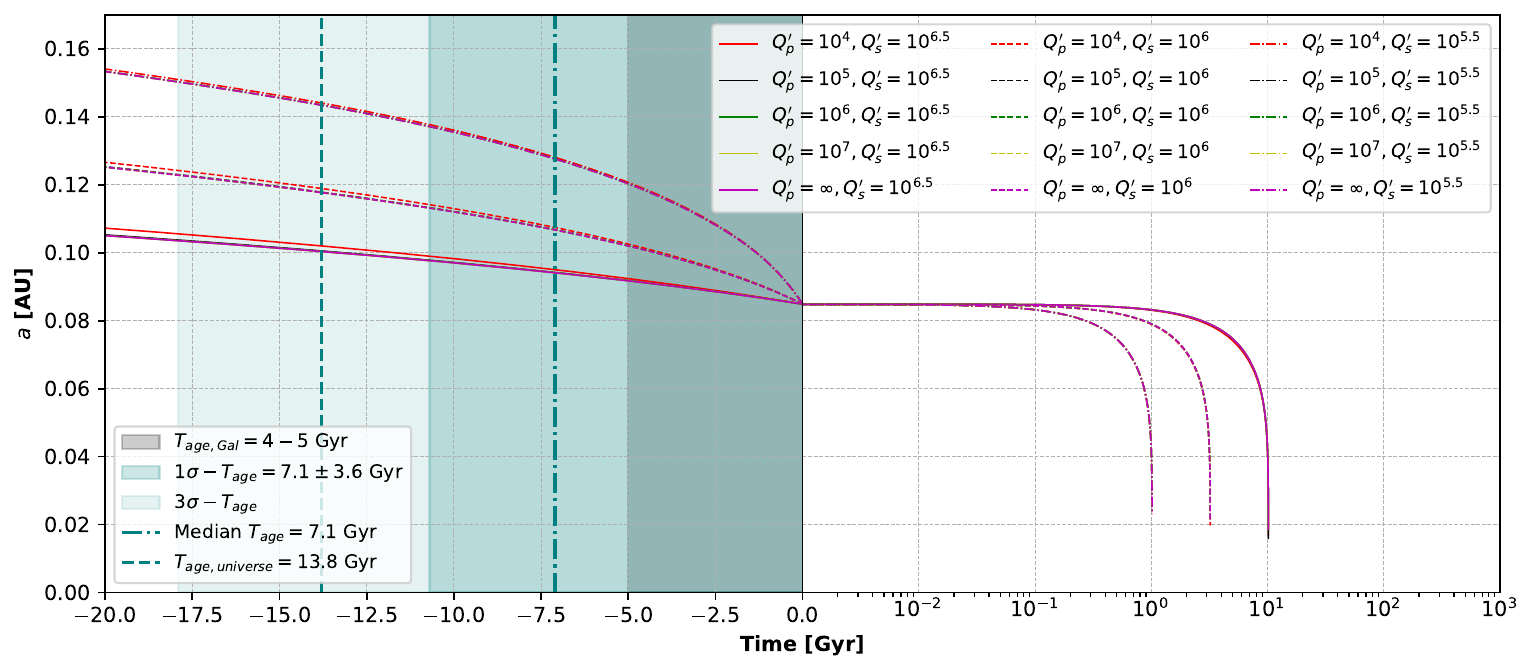}
    \caption{Backward and forward evolution of orbital separation for different $Q_\star'$ and $Q_p'$ values based on the formulation in \citet{Jackson+2009} where the forward evolution is done for $\sim10^3$ Gyr and backward evolution is extrapolated till $\sim20$ Gyr. The grey shaded region represents the age estimate obtained from Galactic kinematics. The shaded teal regions represent the $1\sigma$ and $3\sigma$ regions for the age estimated using \texttt{EXOFASTv2} for decreasing opacity, respectively, and the vertical dash-dotted line represents the median age. For reference, the age of the universe is also plotted as a dashed teal line.}
    \label{fig:a-Qs-full}
\end{figure*}
\begin{figure*}[t!]
    \centering
    \includegraphics[width=\linewidth]{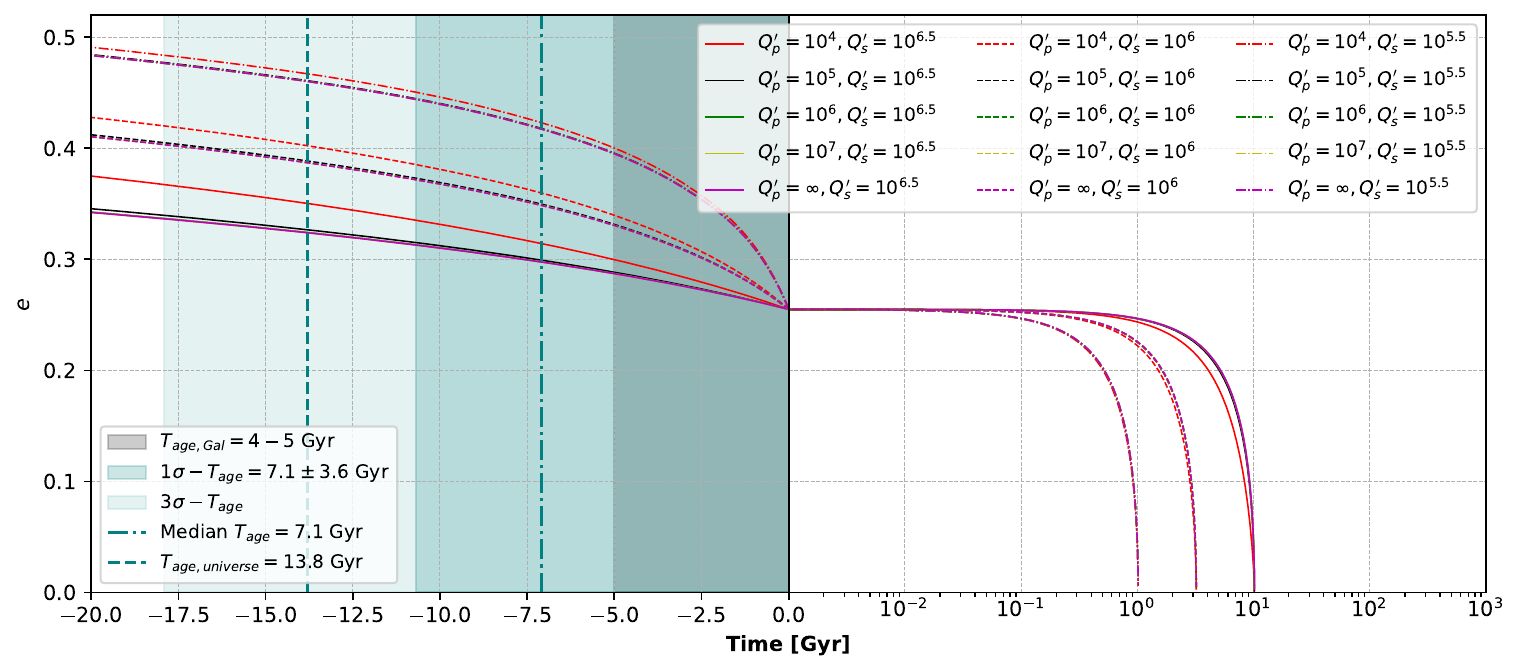}
    \caption{Backward and forward evolution of orbital eccentricity for different $Q_\star'$ and $Q_p'$ values based on the formulation in \citet{Jackson+2009} where the forward evolution is done for $\sim10^3$ Gyr and backward evolution is extrapolated till $\sim20$ Gyr. The grey shaded region represents the age estimate obtained from Galactic kinematics. The shaded teal regions represent the $1\sigma$ and $3\sigma$ regions for the age estimated using \texttt{EXOFASTv2} for decreasing opacity, respectively, and the vertical dash-dotted line represents the median age. For reference, the age of the universe is also plotted as a dashed teal line.}
    \label{fig:e-Qs-full}
\end{figure*}

\begin{figure}[ht!]
\centering
\includegraphics[width=1.0\columnwidth]{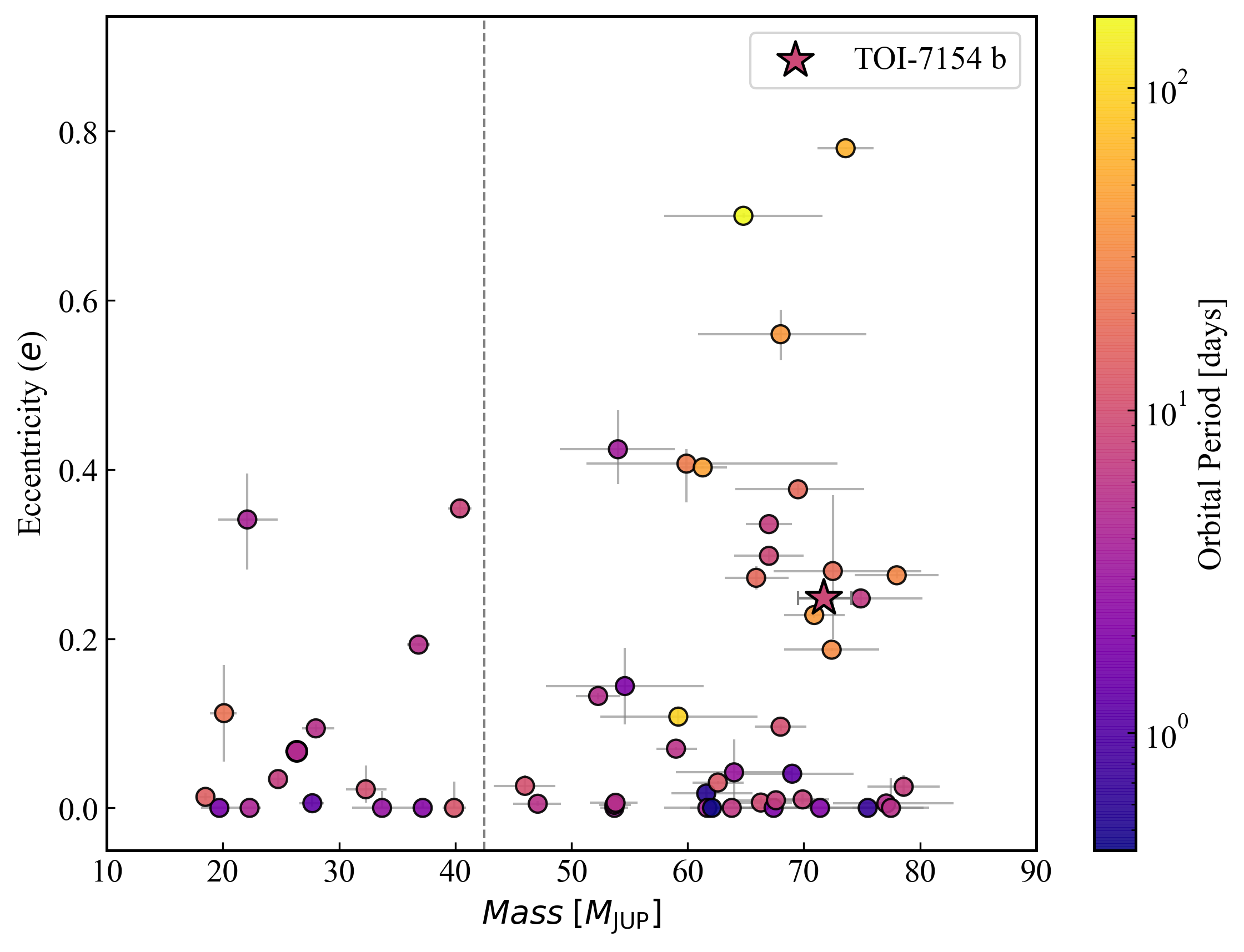}
\caption{{Mass--eccentricity distribution of 56 transiting brown dwarfs, including TOI-6884b (Khandelwal et al., under review) and TOI-7154b (star symbol). The sample of transiting brown dwarfs compiled from the \texttt{TEPCAT} database (see \citet{tepcat} and references therein) and represented by points colored according to their orbital period. The dashed vertical line marks the transition region at $\sim$$42.5\,M_{\mathrm J}$ as per \citet{ma2014}.}}
\label{fig:eccdist}
\end{figure}

Figures \ref{fig:a-Qs-full} and \ref{fig:e-Qs-full} show the evolution of orbital separation and eccentricity, respectively, for TOI-7154 system where the forward evolution is done for $\sim10^3$ Gyr and backward evolution is done till $\sim20$ Gyr. To cover all possible outcomes, we varied $Q_p'$ from $10^4$ to $\infty$ and is represented by varying colors in the plots, while the solid, dashed and dash-dotted lines represent $Q_\star'=10^{6.5}$,  $Q_\star'=10^6$ and $Q_\star'=10^{5.5}$, respectively. We can clearly see that all the colors blend-in for different values of $Q_p'$, except for the case of $Q_p'=10^4$ which shows slight deviation for higher $Q_\star'$ value but gets closer to the rest for smaller $Q_\star'$ value indicating the significant influence of stellar tides in governing the system dynamics. For $Q_\star'=10^{5.5}$, the system will circularise in $\sim 1$ Gyr regardless the impact of planetary tides, which is significantly smaller than what one gets using equation (\ref{eq:tidal4}) (refer Table \ref{tab:tidal1}). This clearly shows that if we neglect the presence of stellar tides, the circularisation timescales obtained for systems with massive planetary or brown-dwarf companions are significantly overestimated. {Since the tidal circularisation timescale is strongly dependent on $Q_\star'$ for TOI-7154 system, we can therefore give an upper bound of $Q_\star'\lesssim10^6$ for the system based on its present state as $\tau_{\text{circ}}\gtrsim T_{\text{age}}$ for any system (considering the age estimated from Galactic kinematics for this system).} It should be noted that the above analysis is done purely from a dynamical point of view conditioned on the present state of the system.

\subsection{Formation scenario}

{TOI-7154b, with a mass of $71.7^{+2.4}_{-2.2}~M_{\mathrm{J}}$, is likely to have formed via star-like processes, opposed to the mechanisms for giant planet formation.} In such a case, its evolution can be compared to the formation of close-in stellar binaries. Most binaries are expected to form through fragmentation \citep[e.g.,][]{bate1995, kroupa1995, kratter2006, clarke2009, Moe2017} at separations greater than 10 AU \citep{boss1986,bate1998}. Close binaries are, therefore, thought to originate at wider separations and subsequently migrate inward. {In environments with high density of stellar sources, tidal capture \citep{Press1977} and dynamical interactions \citep{Hurley2007, Sollima2008} can drive this migration, while in less dense environments, interactions with a third companion may play a role.} For example, the von Zeipel–Lidov-Kozai (vZLK) migration can excite eccentricities in close binaries, with subsequent tidal dissipation shrinking the orbit \citep{Fabrycky2007, Naoz2014}. {However, the migration of a brown dwarf due to secular perturbations from an outer companion depends on the orbital properties of the system and the mass of the companion.} {We performed an ensemble of vZLK-simulations, details of which are presented in Appendix \ref{AppA}, and we find that a migration scenario requires a companion with a mass greater than $\sim0.5\,M_{\odot}$ and a semi-major axis between $\sim500\,\text{AU}$ and $\sim1500\,\text{AU}$ for $Q_\star=10^6$. However, such a nearby companion is ruled out by observations, making this mechanism unlikely. This is also suggestive by the Gaia Renormalised Unit Weighted Error (RUWE) value of $1.04$ \citep{gdr3}. It is, therefore, more likely that the system fragmented at a wide separation and subsequently migrated inward to its current location \citep[e.g.,][]{Bate2002}. It has been shown that many features of the stellar binary properties including the brown dwarf desert can be reproduced by this mechanism \citep{Tokovinin2020}.}

{Therefore, the most promising formation channels for the TOI-7154 system remain stellar-like fragmentation processes. These processes can occur via either turbulent fragmentation due to density perturbations in molecular cores \citep{Kuffmeier+2019}, producing bound systems with no preferential spin-orbit obliquity; or through disk-driven fragmentation processes in rotating molecular clouds \citep{Bodenheimer+1980}, producing moderately eccentric systems with natural spin-orbit alignment \citep{LaughlinBodenheimer1994, Hale1994}. Since we do not have any measurements of spin-orbit obliquity for TOI-7154 system, it is difficult to comment on the exact formation scenario amongst these two. Additionally, gravitational capture can also lead to the formation of these short-period BDs around Sun-like stars, with, however, a very small probability of occurrence \citep{dos-Santos+2024}. This is also coherent with the formation scenarios suggested for two other massive close-in BDs with similar host characteristics, TOI-2533b ($M\approx75\,M_{\mathrm J}$, $a\approx 0.07$ AU, and $e\approx0.25$; \citet{dos-Santos+2024}) and EPIC-212036875b ($M\approx51\,M_{\mathrm J}$, $a\approx0.06$ AU, and $e\approx0.13$; \citet{Persson+2019}), where stellar-like formation mechanisms is proposed.}

{The eccentricity distribution of BDs has been used to probe their origins; \citet{ma2014} identified a transition near $\sim42.5\,M_{\mathrm J}$, below which objects follow the eccentricity distribution of massive planets, and above which they exhibit a broader distribution resembling that of stellar binaries. Subsequent studies have also broadly supported this trend \citep[e.g.,][]{Grieves2017, Kiefer2021}. In Figure~\ref{fig:eccdist}, we show the mass--eccentricity distribution of the known transiting BDs, color-coded with the orbital period. It is evident that the higher-mass BDs ($\gtrsim42.5\,M_{\mathrm J}$) showcase a broader range of eccentricity, while the BDs with the highest eccentricities are found at longer orbital periods, consistent with more efficient tidal circularization at short orbital periods. TOI-7154b lies in the high-mass regime with moderate eccentricity, which indicates incomplete tidal circularization, as discussed in Section~\ref{sec:tidal1}.}

\subsection{Future follow-up opportunities}

While the present work is focused on the detection and characterization of the BD companion, measurements of the Rossiter--McLaughlin (RM) effect \citep{rossiter1924, mclaughlin1924} or Doppler tomography \citep{colliercameron2010} can provide a powerful means of determining the projected spin-orbit alignment between a transiting companion and its host star. Using the estimated radius of the brown dwarf ($0.827^{+0.040}_{-0.037}\,R_{\mathrm{J}}$), the stellar rotational velocity ($v\sin i_\star = 3.3\pm0.5\,\mathrm{km\,s^{-1}}$), and an impact parameter of $b\approx0.7$, we estimated the expected RM semi-amplitude of TOI-7154b to be $\sim 14\,\mathrm{m/s}$. This is computed using the analytic expression from \citet{triaud2018}:
\[
A_{\rm RM} \simeq \frac{2}{3} D\, v\sin i_\star \sqrt{1 - b^2},
\]
where $D$ is the transit depth. Although the host star is relatively faint ($V\sim12.5\,\text{mag}$), such an RM signal may be detectable with ultra-stable, high-resolution spectrographs such as HARPS \citep{harps}, NEID \citep{neid}, EXPRES \citep{expres}, etc. A measurement of the projected obliquity ($\lambda$) would provide valuable insight into the formation and dynamical evolution of the system, particularly in light of TOI-7154b's moderate orbital eccentricity ($e \approx 0.26$). As presented in the previous discussion, since formation mechanisms consistent with fragmentation processes are more likely for TOI-7154b, a determination of the spin-orbit obliquity would serve as a key diagnostic for the viability of the presently conceived formation scenario and whether the system has retained its primordial geometry or undergone later dynamical evolution and tidal re-alignment that has reshaped the system from its primordial configuration.


\section{Summary}\label{sec:summary} 

We report the discovery and detailed characterisation of a high-mass brown dwarf transiting the metal-rich G-type main-sequence star TOI-7154. The system was initially flagged as a transit candidate by NASA’s \textit{TESS} mission, and subsequent high-precision radial velocity measurements obtained with the PARAS-2 and the TRES spectrograph confirmed the presence of a massive sub-stellar companion. Our joint analysis of photometric and spectroscopic data yields well-constrained stellar parameters for TOI-7154 -- iron abundance $\mathrm{[Fe/H]} = 0.154^{+0.077}_{-0.075}\,\text{dex}$, effective temperature $T_{\mathrm{eff}} = 5564^{+100}_{-110}\,\text{K}$, stellar mass $M_\star = 0.939^{+0.047}_{-0.043}\,M_{\odot}$, stellar radius $R_\star = 0.949^{+0.032}_{-0.030}\,R_{\odot}$, surface gravity $\log g = 4.456^{+0.036}_{-0.036}$, and an inferred age of $7.2^{+3.9}_{-3.6}\,\text{Gyr}$. The transiting companion, TOI-7154b, is determined to have a mass of $M_{b} = 71.7^{+2.4}_{-2.2}\,M_{\mathrm{J}}$, a radius of $R_{b} = 0.827^{+0.040}_{-0.037}\,R_{\mathrm{J}}$, and an orbital eccentricity of $e = 0.2482 \pm 0.0024$. With a mass near the hydrogen-burning threshold, TOI-7154b resides in the transitional regime separating BDs from the low-mass stars. Its physical and orbital properties make it an important empirical reference point for testing models of substellar structure and evolution. Given its high mass, significant orbital eccentricity, and location within the sparsely populated \textit{brown dwarf desert}, TOI-7154b represents a desirable target for future multi-wavelength and multi-instrument follow-up observations. Such efforts will help refine its internal and atmospheric properties, while also offering insights into the formation pathways, dynamical histories, and tidal evolution of massive sub-stellar companions.


\begin{acknowledgments}

We gratefully acknowledge the support from PRL-DOS (Department of Space, Government of India) and the Director of PRL for funding the PARAS-2 spectrograph. We are also thankful to the staff at the Mount Abu Observatory for their invaluable assistance during the observations. RS is grateful to Jason Eastman for helpful discussions regarding EXOFASTv2.
CD acknowledges the Param Vikram-1000 High Performance Computing Cluster of the Physical Research Laboratory. HGB acknowledges Cristobal Petrovich from Indiana University, Bloomington, for useful discussions. The work of B.S.S. was conducted under the state assignment of Lomonosov Moscow State University. This research made use of the SIMBAD database and the VizieR catalog access tool, both operated by CDS in Strasbourg, France. We also utilized the Exoplanet Follow-up Observation Program (ExoFOP; DOI: 10.26134/ExoFOP5), managed by the California Institute of Technology under contract with NASA as part of the Exoplanet Exploration Program. Furthermore, this study incorporates data collected by the TESS mission, obtained from the MAST data archive (DOI: \dataset[10.17909/ergc-ph96]{http://dx.doi.org/10.17909/ergc-ph96}) at the Space Telescope Science Institute (STScI), as well as data from the \texttt{TEPCAT} and NASA Exoplanet Archive catalogs. The authors thank the anonymous referee for their valuable suggestions, which enhanced the quality of the paper.

\end{acknowledgments}


\bibliographystyle{aasjournalv7}
\bibliography{toi7154}

\newpage
\providecommand{\bjdtdb}{\ensuremath{\rm {BJD_{TDB}}}}
\providecommand{\tjdtdb}{\ensuremath{\rm {TJD_{TDB}}}}
\providecommand{\feh}{\ensuremath{\left[{\rm Fe}/{\rm H}\right]}}
\providecommand{\teff}{\ensuremath{T_{\rm eff}}}
\providecommand{\teq}{\ensuremath{T_{\rm eq}}}
\providecommand{\ecosw}{\ensuremath{e\cos{\omega_*}}}
\providecommand{\esinw}{\ensuremath{e\sin{\omega_*}}}
\providecommand{\msun}{\ensuremath{\,M_\Sun}}
\providecommand{\rsun}{\ensuremath{\,R_\Sun}}
\providecommand{\lsun}{\ensuremath{\,L_\Sun}}
\providecommand{\mj}{\ensuremath{\,M_{\rm J}}}
\providecommand{\rj}{\ensuremath{\,R_{\rm J}}}
\providecommand{\me}{\ensuremath{\,M_{\rm E}}}
\providecommand{\re}{\ensuremath{\,R_{\rm E}}}
\providecommand{\fave}{\langle F \rangle}
\providecommand{\fluxcgs}{10$^9$ erg s$^{-1}$ cm$^{-2}$}
\onecolumngrid
{\fontsize{9}{12}\selectfont
\renewcommand{\arraystretch}{1.12}
\setlength{\tabcolsep}{6pt} 
\begin{longtable}{lllll}
\caption{Summary of EXOFASTv2 fitted and derived parameters with 68\% confidence interval for TOI-7154 system.}
\label{tab:planet_table} \\

\hline
Parameter & Description & Value \\
\hline
\endfirsthead

\hline
\multicolumn{3}{c}{{\bfseries Table \thetable\ (continued)}} \\
\hline
Parameter & Description & Value \\
\hline
\endhead

\hline \multicolumn{3}{r}{{Continued on next page}} \\
\endfoot

\hline
\endlastfoot

\multicolumn{2}{l}{Stellar Parameters:}\\
~~~~$M_*$\dotfill &Mass (\msun)\dotfill &$0.939^{+0.047}_{-0.043}$\\
~~~~$R_*$\dotfill &Radius (\rsun)\dotfill &$0.949^{+0.032}_{-0.030}$\\
~~~~$R_{*,SED}$\dotfill &Radius$^{1}$ (\rsun)\dotfill &$0.9325^{+0.0099}_{-0.0085}$\\
~~~~$L_*$\dotfill &Luminosity (\lsun)\dotfill &$0.782^{+0.036}_{-0.047}$\\
~~~~$\rho_*$\dotfill &Density (cgs)\dotfill &$1.55^{+0.18}_{-0.16}$\\
~~~~$\log{g}$\dotfill &Surface gravity (cgs)\dotfill &$4.456\pm0.036$\\
~~~~$T_{\rm eff}$\dotfill &Effective temperature (K)\dotfill &$5564^{+100}_{-110}$\\
~~~~$T_{\rm eff,SED}$\dotfill &Effective temperature$^{1}$ (K)\dotfill &$5626^{+65}_{-110}$\\
~~~~$[{\rm Fe/H}]$\dotfill &Metallicity (dex)\dotfill &$0.154^{+0.077}_{-0.075}$\\
~~~~$[{\rm Fe/H}]_{0}$\dotfill &Initial Metallicity$^{2}$ \dotfill &$0.162^{+0.073}_{-0.075}$\\
~~~~$\rm Age$\dotfill &Age (Gyr)\dotfill &$7.2^{+3.9}_{-3.6}$\\
~~~~$\rm EEP$\dotfill &Equal Evolutionary Phase$^{3}$ \dotfill &$368^{+29}_{-28}$\\
~~~~$A_V$\dotfill &V-band extinction (mag)\dotfill &$0.173^{+0.052}_{-0.085}$\\
~~~~$\sigma_{\rm SED}$\dotfill &SED photometry error scaling \dotfill &$0.78^{+0.35}_{-0.21}$\\
~~~~$\varpi$\dotfill &Parallax (mas)\dotfill &$3.365\pm0.015$\\
~~~~$d$\dotfill &Distance (pc)\dotfill &$297.2^{+1.4}_{-1.3}$\\
\smallskip\\
\hline
\smallskip\\\multicolumn{2}{l}{Planetary Parameters:}\smallskip\\
~~~~$P$\dotfill &Period (days)\dotfill &$8.860073\pm0.000029$\\
~~~~$R_b$\dotfill &Radius (\rj)\dotfill &$0.827^{+0.040}_{-0.037}$\\
~~~~$M_b$\dotfill &Mass (\mj)\dotfill &$71.7^{+2.4}_{-2.2}$\\
~~~~$T_C$\dotfill &Observed Time of conjunction$^{4}$ (\bjdtdb)\dotfill &$2460464.0510^{+0.0021}_{-0.0022}$\\
~~~~$T_C$\dotfill &Model Time of conjunction$^{4,5}$ (\tjdtdb)\dotfill &$2460464.0505^{+0.0021}_{-0.0022}$\\
~~~~$T_T$\dotfill &Model time of min proj sep$^{5,6,7}$ (\tjdtdb)\dotfill &$2459923.5866\pm0.0012$\\
~~~~$T_0$\dotfill &Obs time of min proj sep$^{6,8,9}$ (\bjdtdb)\dotfill &$2459923.5870\pm0.0012$\\
~~~~$a$\dotfill &Semi-major axis (AU)\dotfill &$0.0840^{+0.0014}_{-0.0013}$\\
~~~~$i$\dotfill &Inclination (Degrees)\dotfill &$87.66\pm0.17$\\
~~~~$e$\dotfill &Eccentricity \dotfill &$0.2482\pm0.0024$\\
~~~~$\omega_*$\dotfill &Arg of periastron (Degrees)\dotfill &$179.30^{+0.88}_{-1.1}$\\
~~~~$T_{\rm eq}$\dotfill &Equilibrium temp$^{10}$ (K)\dotfill &$902^{+12}_{-13}$\\
~~~~$K$\dotfill &RV semi-amplitude (m/s)\dotfill &$7232^{+35}_{-33}$\\
~~~~$R_b/R_*$\dotfill &Radius of planet in stellar radii \dotfill &$0.0896\pm0.0022$\\
~~~~$a/R_*$\dotfill &Semi-major axis in stellar radii \dotfill &$19.03^{+0.71}_{-0.69}$\\
~~~~$\delta$\dotfill &$\left(R_P/R_*\right)^2$ \dotfill &$0.00803^{+0.00041}_{-0.00040}$\\
~~~~$\tau$\dotfill &In/egress transit duration (days)\dotfill &$0.0188^{+0.0019}_{-0.0017}$\\
~~~~$T_{14}$\dotfill &Total transit duration (days)\dotfill &$0.1164^{+0.0027}_{-0.0026}$\\
~~~~$T_{\rm FWHM}$\dotfill &FWHM transit duration (days)\dotfill &$0.0974^{+0.0031}_{-0.0030}$\\
~~~~$b$\dotfill &Transit impact parameter \dotfill &$0.726^{+0.028}_{-0.032}$\\
~~~~$\rho_b$\dotfill &Density (cgs)\dotfill &$157^{+25}_{-22}$\\
~~~~$logg_P$\dotfill &Surface gravity (cgs)\dotfill &$5.415\pm0.044$\\
~~~~$\Theta$\dotfill &Safronov Number \dotfill &$15.50^{+0.73}_{-0.71}$\\
~~~~$\fave$\dotfill &Incident Flux (\fluxcgs)\dotfill &$0.1414^{+0.0075}_{-0.0081}$\\
~~~~$T_P$\dotfill &Time of Periastron (\tjdtdb)\dotfill &$2460456.697^{+0.021}_{-0.025}$\\
~~~~$T_A$\dotfill &Time of asc node (\tjdtdb)\dotfill &$2460461.155^{+0.020}_{-0.016}$\\
~~~~$T_D$\dotfill &Time of desc node (\tjdtdb)\dotfill &$2460456.7072^{+0.0092}_{-0.011}$\\
~~~~$e\cos{\omega_*}$\dotfill & \dotfill &$-0.2481\pm0.0024$\\
~~~~$e\sin{\omega_*}$\dotfill & \dotfill &$0.0030^{+0.0047}_{-0.0038}$\\
~~~~$M_b\sin i$\dotfill &Minimum mass (\mj)\dotfill &$71.7^{+2.4}_{-2.2}$\\
~~~~$M_b/M_*$\dotfill &Mass ratio \dotfill &$0.0730\pm0.0013$\\
~~~~$d/R_*$\dotfill &Separation at mid transit \dotfill &$17.80^{+0.67}_{-0.65}$\\
\smallskip\\\multicolumn{2}{l}{Wavelength Parameters:}&TESS\smallskip\\
~~~~$u_{1}$\dotfill &Linear limb-darkening coeff \dotfill &$0.342\pm0.031$\\
~~~~$u_{2}$\dotfill &Quadratic limb-darkening coeff \dotfill &$0.256^{+0.027}_{-0.026}$\\
\smallskip\\\multicolumn{2}{l}{Telescope Parameters:}&PARAS-2&TRES\smallskip\\
~~~~$\gamma_{\rm rel}$\dotfill &Relative RV Offset (m/s)\dotfill &$558^{+95}_{-96}$&$-5462\pm21$\\
~~~~$\sigma_J$\dotfill &RV Jitter (m/s)\dotfill &$411^{+93}_{-70}$&$50^{+45}_{-22}$\\
~~~~$\sigma_J^2$\dotfill &RV Jitter Variance \dotfill &$1.69^{+0.85}_{-0.53} \times 10^{5}$&$2500^{+6600}_{-1700}$\\
\smallskip\\\multicolumn{5}{l}{Transit Parameters:}\smallskip\\
 & 2020-06-08 (TESS) & 2022-05-09 (TESS) & 2022-06-03 (TESS) & 2025-04-12 (TESS) \smallskip\\
~~~~Added Variance ($\sigma^{2}$) &$2.73^{+0.32}_{-0.31} \times 10^{-6}$ & $3.51^{+0.35}_{-0.33} \times 10^{-6}$ & $-2.7^{+2.3}_{-2.2} \times 10^{-7}$ & $1.849^{+0.074}_{-0.073} \times 10^{-5}$\\
~~~~Baseline flux ($F_0$) &$1.000147\pm0.000080$&$1.000035^{+0.000070}_{-0.000068}$&$0.999964\pm0.000054$&$1.000145^{+0.000082}_{-0.000083}$\\
\noalign{\smallskip}
\hline
\noalign{\smallskip}
\end{longtable}
}
\newpage
\begin{table*}[h]
\centering
\caption{RV measurements of TOI-7154} 
\label{tab:rv_table}
\begin{tabular}{ccccc}
\hline
\hline
\noalign{\smallskip}
BJD$_{TDB}$& Relative-RV & $\sigma$-RV & Exposure Time  & Instrument \\
Days & km s$^{-1}$ & m s$^{-1}$ & sec & -\\
\noalign{\smallskip}
\hline
\noalign{\smallskip}
2460754.499058 &   5.11 &   0.16 & 3600 & PARAS-2\\
2460757.435165 &  -7.07 &   0.05 & 3600 & PARAS-2\\
2460758.475037 &  -7.31 &   0.05 & 3600 & PARAS-2\\
2460759.427958 &  -1.78 &   0.04 & 3600 & PARAS-2\\
2460760.484914 &   3.15 &   0.03 & 3600 & PARAS-2\\
2460761.427243 &   5.82 &   0.04 & 3600 & PARAS-2\\
2460762.433187 &   7.16 &   0.60 & 3600 & PARAS-2\\
2460763.469435 &   5.62 &   0.10 & 3600 & PARAS-2\\
2460764.414171 &   2.20 &   0.05 & 3600 & PARAS-2\\
2460765.404731 &  -1.90 &   0.06 & 3600 & PARAS-2\\
2460766.426127 &  -7.77 &   0.06 & 3600 & PARAS-2\\
2460767.414766 &  -6.32 &   0.05 & 3600 & PARAS-2\\
2460768.434135 &  -0.88 &   0.04 & 3600 & PARAS-2\\
2460770.401918 &   5.62 &   0.03 & 3600 & PARAS-2\\
2460771.417466 &   5.86 &   0.10 & 3600 & PARAS-2\\
2460772.421867 &   4.93 &   0.12 & 3600 & PARAS-2\\
2460782.457916 &   1.55 &   0.07 & 3600 & PARAS-2\\
2460784.224408 &  -8.50 &   0.08 & 3600 & PARAS-2\\
2460785.249346 &  -5.22 &   0.11 & 3600 & PARAS-2\\
2460786.353444 &  -0.27 &   0.13 & 3600 & PARAS-2\\
2460926.7039 &  -13.62 &   0.03 & 2520 & TRES\\
2460948.6462 &   -0.04 &   0.02 & 2280  & TRES\\
2460950.6443 &   -3.88 &   0.02 & 1950 & TRES\\
2460954.6113 &   -6.14 &   0.03 & 1600 & TRES\\
2460956.6160 &   -0.41 &   0.02 & 2150 & TRES\\
2460973.5894 &   -1.54 &   0.02 & 2400 & TRES\\
2460977.5880 &   -5.41 &   0.02 & 1650 & TRES\\
2460978.5899 &  -11.26 &   0.03 & 1740 & TRES\\
\noalign{\smallskip}
\hline
\noalign{\smallskip}
\end{tabular}
\end{table*}

\clearpage

\appendix
\counterwithin{figure}{section}
\counterwithin{table}{section}

\section{von Ziepel-Lidov-Kozai simulations}\label{AppA}

We present an ensemble of secular simulations carried out to explore the parameter space in which migration could be possible. The theoretical construct for these simulations is adopted as discussed in \citet{Petrovich2015}, wherein the interaction potential between the inner and outer orbits is expanded up to octupolar order in the ratio of their semi-major axes. We model the tidal dissipation in the inner binary using the equilibrium tide model as per \citet{Eggleton+1998}. We also include apsidal precession due to general relativity and rotational deformation. In addition, we use the empirical stability criteria from \citet{Petrovich2015Stab} to check for the stability of the system. It should be noted that when the inner orbit is very eccentric, the brown dwarf is assumed to be tidally disrupted ($a(1-e)<2.7R_{*}\sim0.013$ AU, see \citet{Guillochon+2011} for details).

\begin{figure*}[h!]
\centering
\includegraphics[width=0.325\textwidth]{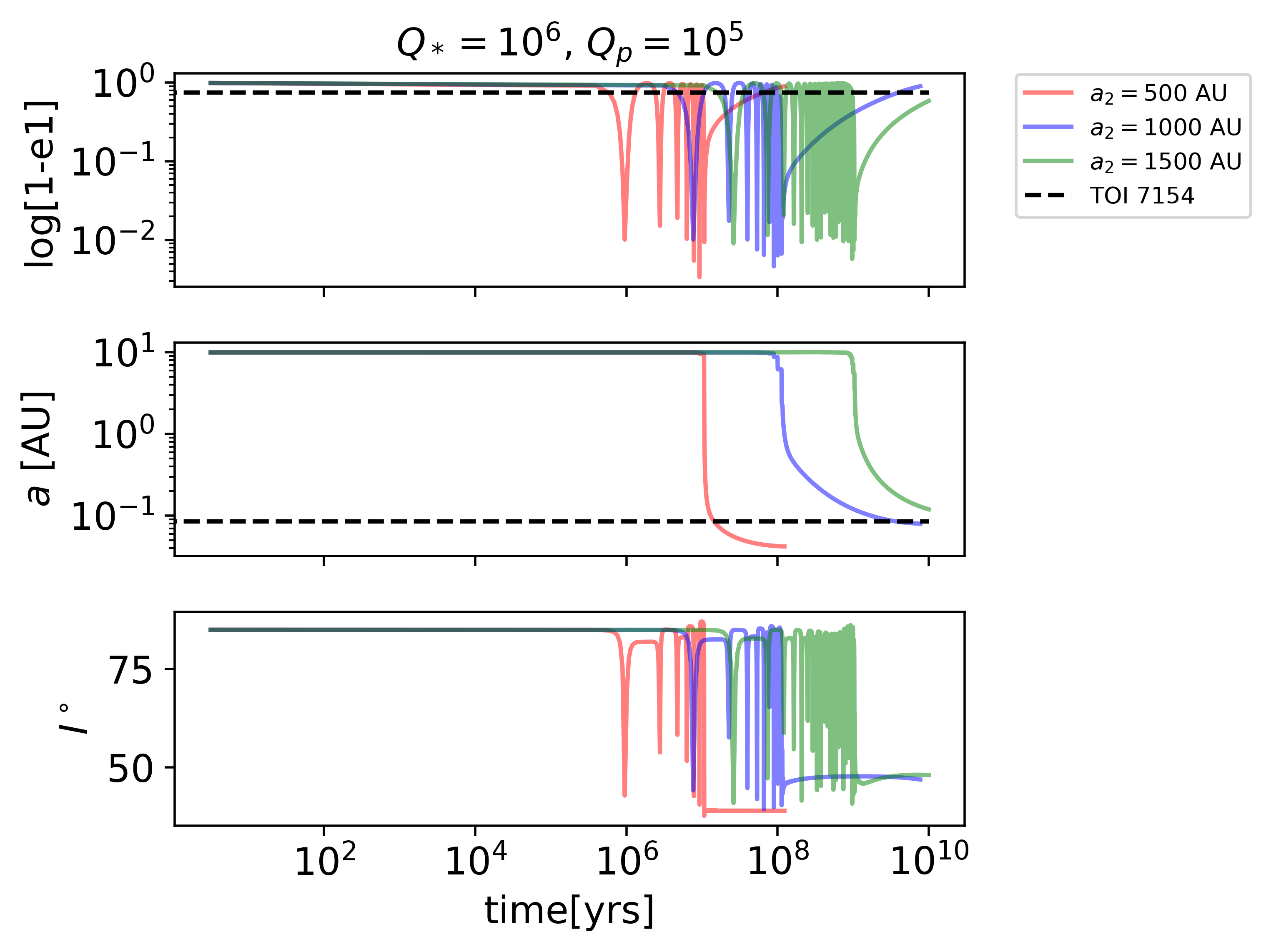}
\includegraphics[width=0.325\textwidth]{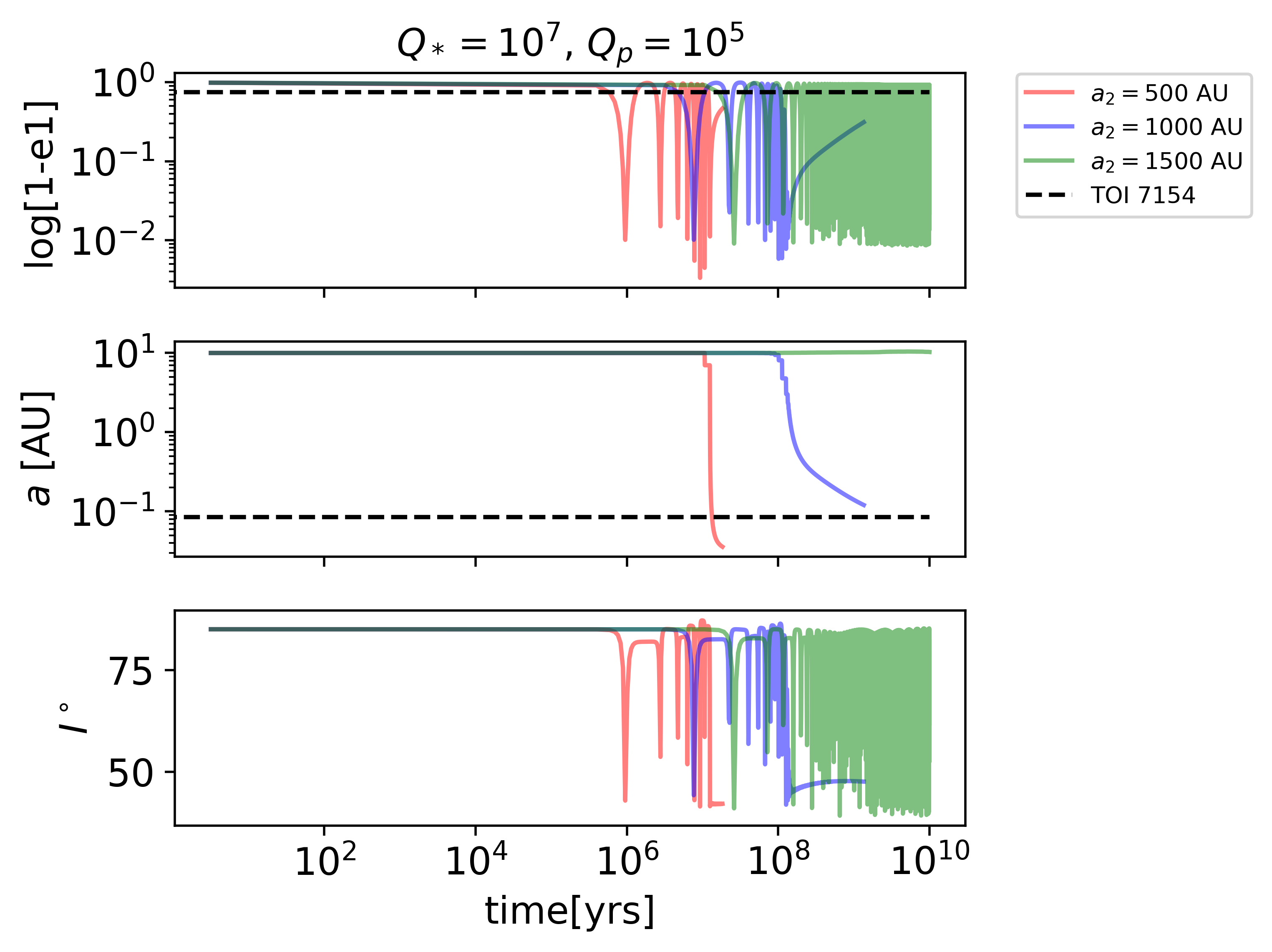}
\includegraphics[width=0.325\textwidth]{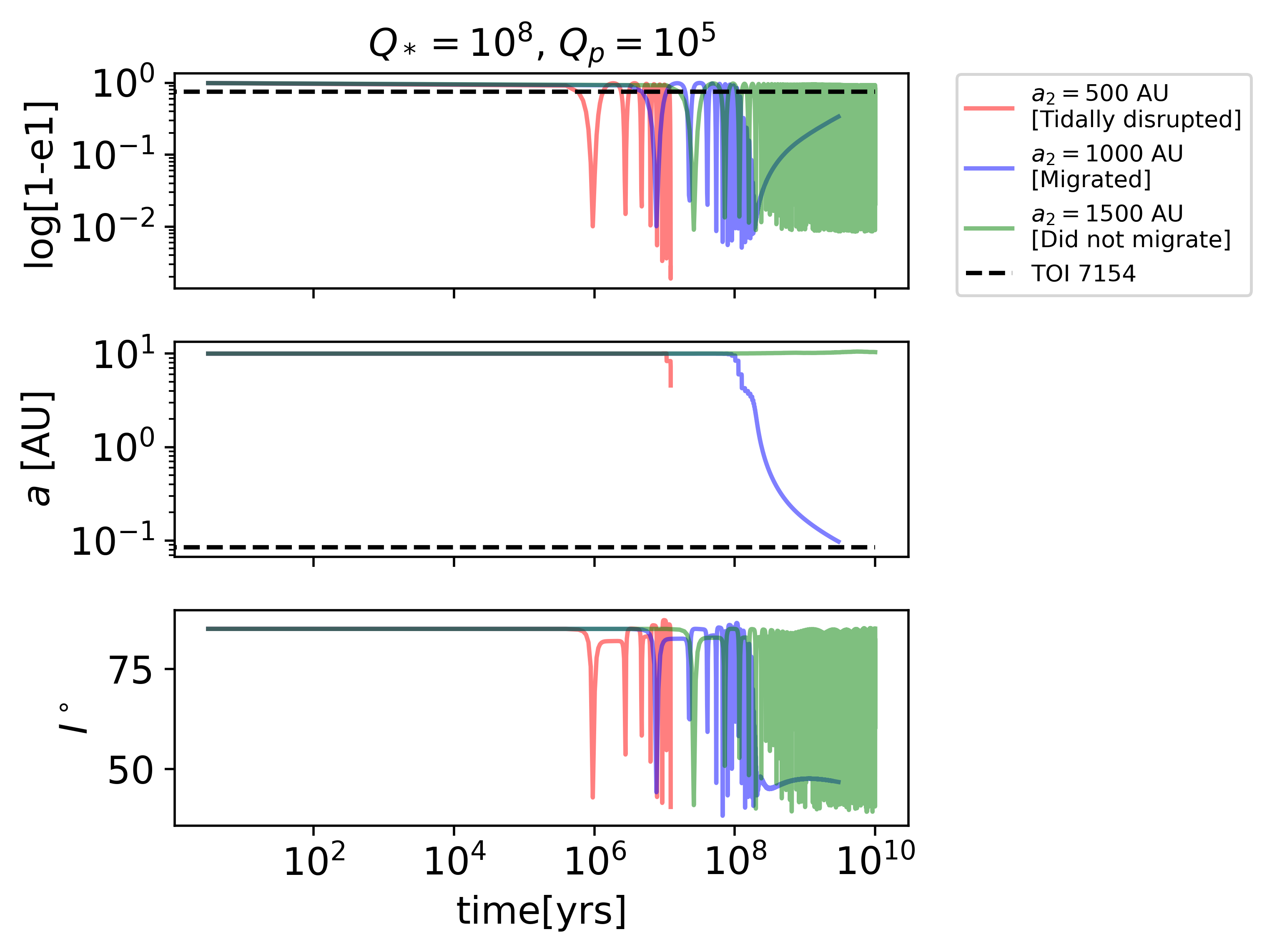}
\caption{The time evolution of eccentricity (top panel), semi-major axis (middle panel) and inclination (bottom panel) of a brown dwarf perturbed by a $1\,M_{\odot}$
companion. The colors show the semi-major axis of the companion, while the dashed black lines mark the observed semi-major axis and eccentricity of TOI-7154b in the middle and top panels, respectively. The brown dwarf depicted in green does not migrate because the companion resides in a wide orbit. In contrast, the brown dwarf shown in red is tidally disrupted when the companion is close-in, as the secular perturbation timescale is short compared to the tidal circularizationn timescale. Finally, the case shown in blue undergoes successful migration. Initially, the brown dwarf is placed in a circular orbit at a distance of $10$ AU from the host star, and with the eccentricity of the companion. The initial mutual inclination between the orbits is set to $85^\circ$. The tidal love number and quality factors of the brown dwarf are $k_{\text{p}}=0.5$ and $Q_{\text{p}'}=10^5$, respectively. The tidal quality factors for the star are $Q_\star' = 10^6$, $Q_\star' = 10^7$ and $Q_\star' = 10^8$ for the left, center, and right panels, respectively.}
\label{fig:tidalsim1}
\end{figure*}
\begin{figure*}[h!]
\centering
\includegraphics[width=0.325\textwidth]{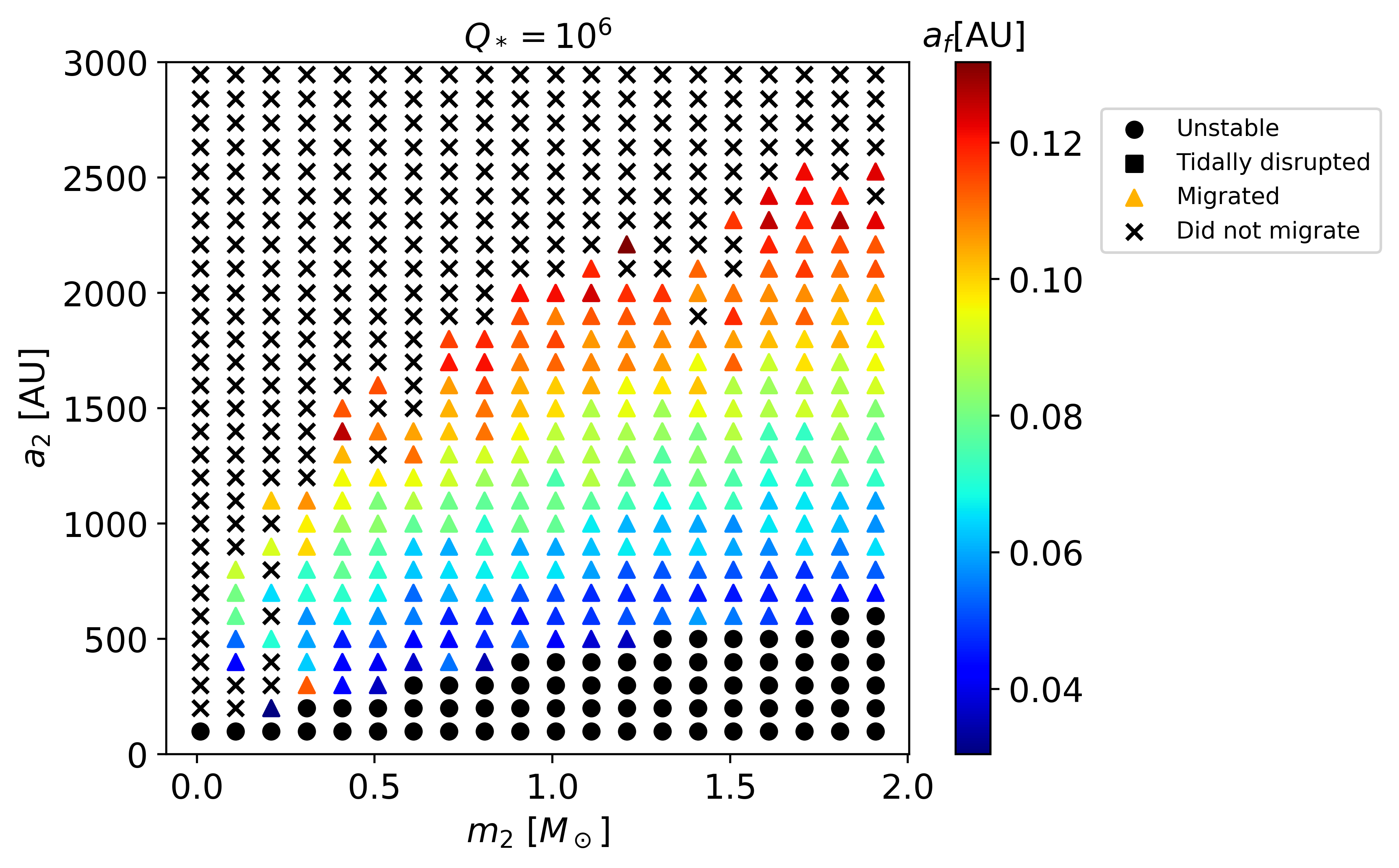}
\includegraphics[width=0.325\textwidth]{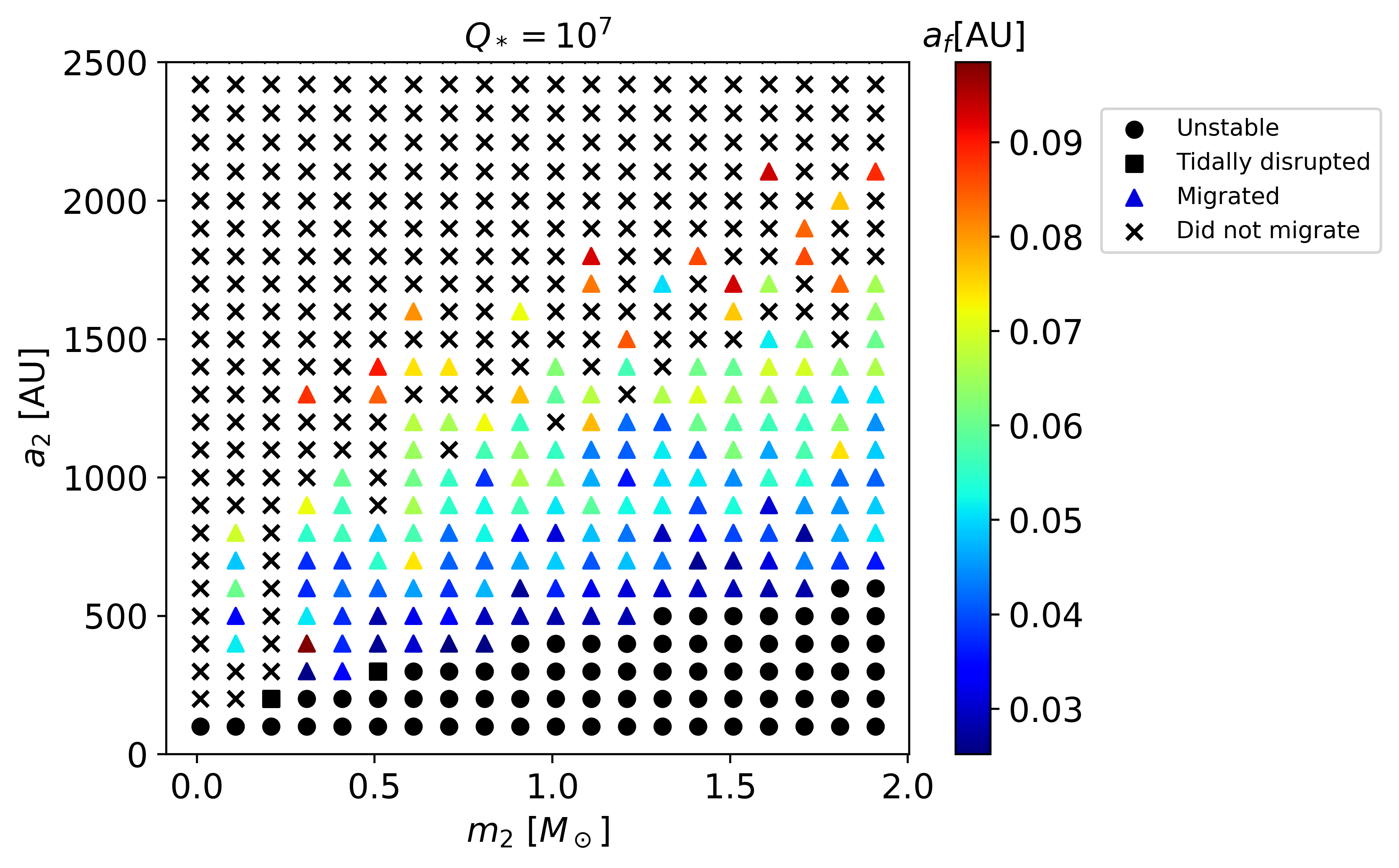}
\includegraphics[width=0.325\textwidth]{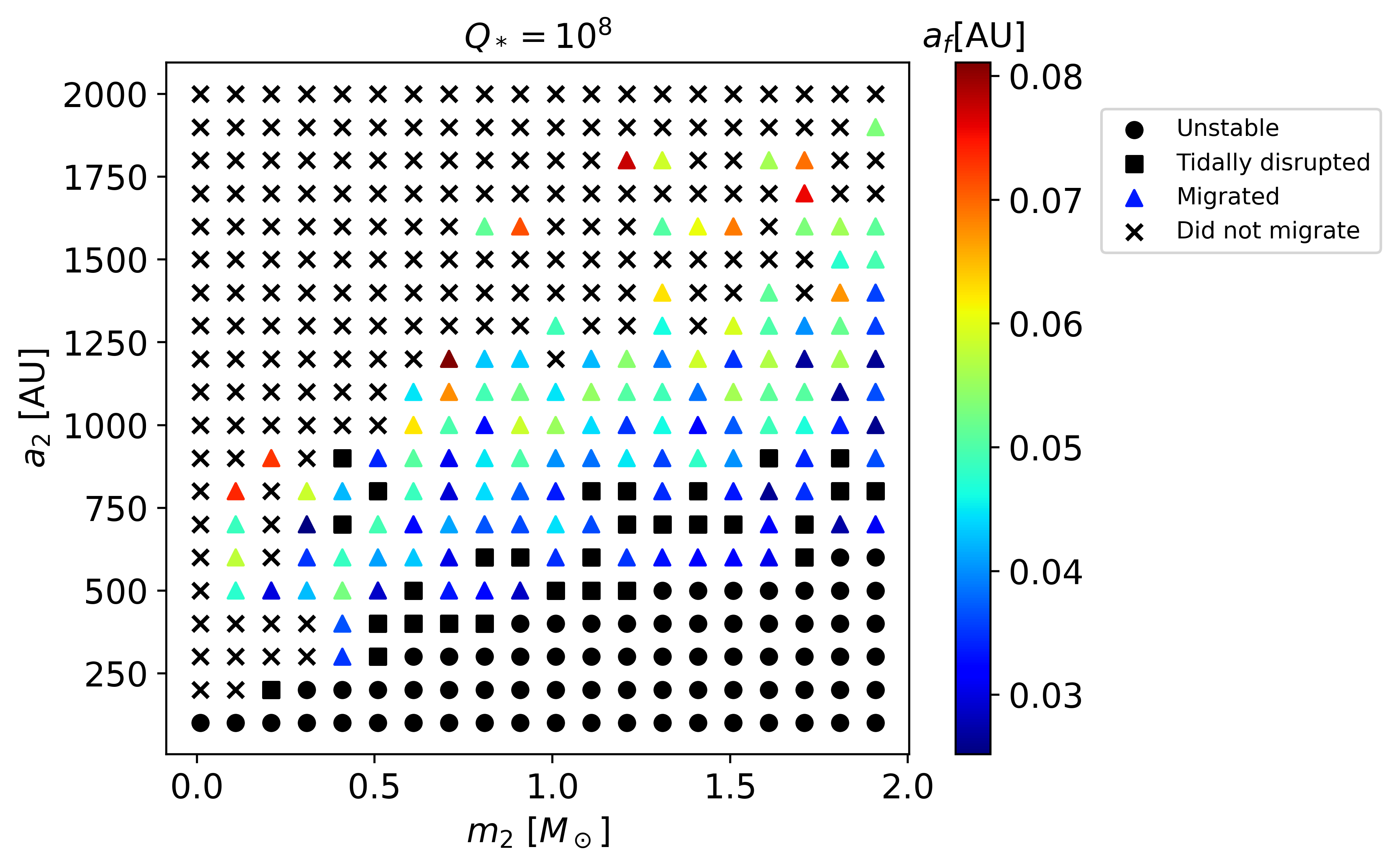}
\caption{The results from our ensemble simulations. The horizontal axis represents the mass of the companion (in $M_{\odot}$), while the vertical axis corresponds to the initial semi-major axis of the outer orbit (in $\text{AU}$).
Configurations that are unstable are marked with black circles, while tidal disruptions are indicated by black squares. Runs in which the brown dwarfs do not migrate are shown with crosses, and those that successfully migrate are represented by triangles. The color scale denotes the final semi-major axis of the brown dwarf. These results demonstrate that close-in companions lead either to instability or tidal disruption, whereas systems with wide-orbit companions do not experience migration. Notably, a solar-mass companion at approximately 1000 AU can facilitate high-eccentricity migration of the brown dwarf. The initial semi-major axis of the brown dwarf is set to 10 AU. The mutual inclination between the inner and the outer orbit is set to $85^\circ$. The love number and tidal quality factors of the brown dwarf are $k_{\text{p}}=0.5$ and $Q_{\text{p}'}=10^5$ respectively. The tidal quality factors for the star are $Q_\star'=10^6$, $Q_\star'=10^7$ and $Q_\star'=10^8$ for the left, center and right panels, respectively.}
\label{fig:tidalsim2}
\end{figure*}

Figure \ref{fig:tidalsim1} shows the time evolution of eccentricity, semi-major axis, and inclination for selected brown dwarfs in the simulated systems for differnet values of $Q_\star'$. The colors represent the semi-major axis of the companion. For $Q_\star'=10^8$, when the companion is far away ($a_2 = 1500\,\text{AU}$), the brown dwarf does not migrate. In contrast, a close-in companion ($a_2 = 500\,\text{AU}$) triggers tidal disruption. Finally, a stellar companion at $1000\,\text{AU}$ can drive the brown dwarf into a close-in orbit within $\sim10^8$ years. The situation is quite different for $Q_\star'=10^6$ where migration happens even for a stellar companion at $1500\,\text{AU}$. Figure \ref{fig:tidalsim2} presents the results of our ensemble simulations in terms of the semi-major axis and mass of the companion for different $Q_\star'$ values, where again the range of vZLK-driven migration reaches as far as $\sim2000\,\text{AU}$ for a solar-mass perturber for $Q_\star'=10^6$. In general, similar to Figure \ref{fig:tidalsim1}, close-in companions either lead to unstable systems or tidal disruptions, while systems with very wide-orbit companions do not experience migration. In the intermediate regime, however, brown dwarfs can migrate close to their host stars. The scales depend on the tidal deformability of the host star and the brown dwarf. It should also be noted that for $Q_\star=10^6$, a larger fraction of the ensemble undergoes successful migration (shown by triangles) compared to $Q_\star=10^8$. This shows that stronger tidal impacts may lead to higher migration probability in a given population ensemble for a system like TOI-7154, as stronger tidal influence leads to faster tidal dissipation and subsequent circularisation leading to migration, before disruption can happen. 

We'd like to highlight that these simulations are not exhaustive and do not cover the full parameter space, for instance, the initial inclination is fixed at $85^\circ$ and a closer-in companion could migrate without disruption if the initial inclination were lower. Fully exploring this broad parameter space would require a more rigorous analysis that is beyond the scope of the current work.

\end{document}